\UseRawInputEncoding
\documentclass[%
 reprint,
superscriptaddress,
 amsmath,amssymb,
 aps,
prb,
]{revtex4-1}

\usepackage{graphicx}% Include figure files
\usepackage{dcolumn}% Align table columns on decimal point
\usepackage{bm}% bold math
\usepackage{amsmath,amssymb}
\usepackage{graphicx}
\usepackage{color}

\usepackage{hyperref}% add hypertext capabilities

\begin{document}
%\setstretch{1.5}

\title{Liquid-assisted laser nanotexturing of silicon: onset of hydrodynamic processes regulated by laser-induced periodic surface structures}

\author{Yulia Borodaenko}
\affiliation{Institute of Automation and Control Processes, Far Eastern Branch, Russian Academy of Science, 5 Radio Str., Vladivostok 690041, Russia}

\author{Dmitriy Pavlov}
\affiliation{Institute of Automation and Control Processes, Far Eastern Branch, Russian Academy of Science, 5 Radio Str., Vladivostok 690041, Russia}

\author{Artem Cherepakhin}
\affiliation{Institute of Automation and Control Processes, Far Eastern Branch, Russian Academy of Science, 5 Radio Str., Vladivostok 690041, Russia}

\author{Eugeny Mitsai}
\affiliation{Institute of Automation and Control Processes, Far Eastern Branch, Russian Academy of Science, 5 Radio Str., Vladivostok 690041, Russia}

\author{Andrei Pilnik}
\affiliation{Pacific Quantum Center, Far Eastern Federal University, Vladivostok, Russia}
\affiliation{Institute of Automation and Control Processes, Far Eastern Branch, Russian Academy of Science, 5 Radio Str., Vladivostok 690041, Russia}

\author{Sergey Syubaev}
\affiliation{Institute of Automation and Control Processes, Far Eastern Branch, Russian Academy of Science, 5 Radio Str., Vladivostok 690041, Russia}
\affiliation{Image Processing Systems Institute of RAS – Branch of the FSRC "Crystallography and Photonics" RAS, Samara, Russia}

\author{Stanislav O. Gurbatov}
\affiliation{Institute of Automation and Control Processes, Far Eastern Branch, Russian Academy of Science, 5 Radio Str., Vladivostok 690041, Russia}
\affiliation{Far Eastern Federal University, Vladivostok, Russia}

\author{Evgeny Modin}
\affiliation{CIC NanoGUNE BRTA, Avda Tolosa 76, 20018 Donostia-San Sebastian, Spain}

\author{Aleksey P. Porfirev}
\affiliation{Image Processing Systems Institute of RAS – Branch of the FSRC "Crystallography and Photonics" RAS, Samara, Russia}

\author{Svetlana N. Khonina}
\affiliation{Image Processing Systems Institute of RAS – Branch of the FSRC "Crystallography and Photonics" RAS, Samara, Russia}

\author{Aleksandr V. Shevlyagin}
\affiliation{Institute of Automation and Control Processes, Far Eastern Branch, Russian Academy of Science, 5 Radio Str., Vladivostok 690041, Russia}

\author{Evgeny L. Gurevich}
\email{gurevich@fh-muenster.de}
\affiliation{Laser Center (LFM), University of Applied Sciences Munster, Stegerwaldstra?e 39, 48565 Steinfurt, Germany}

\author{Aleksandr A. Kuchmizhak}
\email{alex.iacp.dvo@mail.ru}
\affiliation{Institute of Automation and Control Processes, Far Eastern Branch, Russian Academy of Science, 5 Radio Str., Vladivostok 690041, Russia}
\affiliation{Pacific Quantum Center, Far Eastern Federal University, Vladivostok, Russia}

\begin{abstract}
Ultrashort laser exposure can push materials to transient highly nonequilibrium states driving their subsequent morphology self-organization stimulated by excited electromagnetic and hydrodynamic processes. Laser processing parameters regulate these processes defining the resulting surface nano-morphologies demanding for diverse applications. Here, upon systematic studies of femtosecond-laser processing of monocrystalline Si in oxidation-preventing methanol, we showed that the electromagnetic processes dominating at initial steps of the progressive morphology evolution define the onset of the hydrodynamic processes and resulting morphology upon subsequent multi-pulse exposure. In particular, under promoted exposure quasi-regular subwavelength laser-induced periodic surface structures (LIPSSs) were justified to evolve through the template-assisted development of the Rayleigh-Plateau hydrodynamic instability in the molten ridges forming quasi-regular surface patterns with a supra-wavelength periodicity and preferential alignment along polarization direction of the incident light. Subsequent exposure promotes fusion-assisted morphology rearrangement resulting in a spiky surface with a random orientation, yet constant inter-structure distance correlated with initial LIPSS periodicity. Along with the insight onto the physical picture driving the morphology evolution and supra-wavelength nanostructure formation, our experiments also demonstrated that the resulting quasi-regular and random spiky morphology can be tailored by the intensity/polarization distribution of the incident laser beam allowing on-demand surface nanotexturing with diverse hierarchical surface morphologies exhibiting reduced reflectivity in the visible and shortwave IR spectral ranges. Finally, we highlighted the practical attractiveness of the suggested approach for improving near-IR photoresponse and expanding operation spectral range of vertical p-n junction Si photo-detector operating under room temperature and zero-bias conditions $via$ single-step annealing-free laser nanopatterning of its surface.
\end{abstract}

\maketitle

\section{Introduction}
Surface micro- and nanostructuring represents an important technological procedure allowing to improve basic characteristics of critical materials or even endow them with new functionalities. Bright examples of such material improvement can be readily found in almost all aspects of modern engineering and device fabrication, where appropriate patterning allows to control electrical, optical, biological, tribological and wetting properties of the processed surfaces. Direct laser processing of materials using pulsed radiation represents non-lithographic straightforward technique exhibiting continuously growing importance for both scientific and industrial communities. Such interest is largely boosted by the development of the laser market offering cheaper, high-speed and stable pulsed sources, more precise beam shaping and scanning systems as well as the growth of fundamental understanding of laser-matter interaction $via$ development of computer-aided simulation tools and analysis techniques based on machine learning \cite{sugioka2014ultrafast,vorobyev2013direct,wei2020overview,stratakis2020laser,dalloz2022anti,porfirev2023light,brandao2023learning}.

Apart from a common direct patterning of the surface with a tightly focused laser beam along the computer-defined trajectory, the formation of surface features $via$ laser-driven self-organization phenomena has gained significant research interest during past decades owing to practically attractive simplicity, fabrication scalability and ability of such an approach to create diverse regular or random surface morphologies with characteristic nanoscale footprints \cite{stoian2020advances,stoian2023ultrafast}. Laser-induced periodic surface structures (LIPSSs; also referred to as ripples) discovered in 1965 \cite{birnbaum1965semiconductor} shortly after intensification of the studies related to laser-matter interaction  represent remarkable example of such self-organization driven by pulsed (or even CW \cite{nemanich1983aligned}) laser radiation on the surface of practically all types of materials \cite{buividas2014surface,bonse2016laser,bonse2020maxwell}. LIPSSs usually appear as a (quasi)regular grating-type surface morphology aligned perpendicularly or along the polarization vector of the incident radiation depending on the dominating contribution of the effects driven self-organization. In most cases, these effects have "electromagnetic" nature explaining the periodical surface ordering $via$ interference of the incident radiation with the surface/scattered waves that yield a characteristic LIPSSs periodicity $\Lambda$ close to the laser wavelength $\lambda$ used. Specific experimental conditions (such as texturing in liquids) or processing regimes driving the laser-irradiated material into the nonequlibrium state and facilitating interference of the counter-propagating surface waves allow to shrink the grating periodicity down to deep subwavelength values ($\Lambda<<\lambda$; high spatial frequency LIPSSs). Supra-wavelength surface patterns with the periodicity $\Lambda>\lambda$, which formation indeed falls outside electromagnetic scenario, were also reported on the surface of diverse material manifesting themselves as universal phenomenon and unveiling complexity of the laser-driven self-organization processes that are yet far from complete understanding \cite{tsibidis2016convection,xu2019periodic,zhang2021irregular}.

Monocrystalline Si (c-Si), a critically important semiconductor widely applied in variety of optoelectronic and nanophotonic devices \cite{green2021solar,baranov2017all}, was widely studied in the context of LIPSSs formation revealing the ability to produce on its surface both high spatial frequency LIPSSs \cite{le2005sub,shen2008high,le2011generation,straub2012periodic,hamad2014femtosecond,meng2017dual,yiannakou2017cell,zhang2019hierarchical,kesaev2021nanopatterned,borodaenko2021deep,borodaenko2022demand} as well as quasi-regular supra-wavelength patterns \cite{ma2014progressive,zayarny2016surface,nivas2017femtosecond,nivas2018direct,allahyari2020formation,nivas2021incident,hu2022ultrafast,kawabata2023two}. The latter type of the structures was generally explained by hydrodynamic processes \cite{tsibidis2015ripples}, yet their ordering along polarization direction and evident correlation between the characteristic periodicity and laser wavelength were established \cite{nivas2018direct} reflecting the lack of complete understanding of the underlying physical processes driven morphology self-organization. In this respect, more systematic studies targeted at unveiling the role of hydrodynamic and electromagnetic processes in formation of supra-wavelength structures are required to achieve full-scale control over the surface morphology crucial for obtaining desired optical properties as well as device performance optimization.

In this paper, we systematically studied fs-laser processing of c-Si immersed in methanol revealing that the electromagnetic processes dominating at initial steps of the progressive morphology evolution define the onset of the hydrodynamic processes and resulting morphology upon subsequent multi-pulse exposure. In particular, under promoted exposure regular subwavelength LIPSSs were shown to evolve through the template-assisted development of the Rayleigh-Plateau (R.-P.) hydrodynamic instability in the molten ridges creating quasi-regular surface patterns with supra-wavelength periodicity and alignment along polarization direction of the incident light. Subsequent exposure promotes fusion-assisted morphology rearrangement resulting in spiky surface with random orientation, yet constant inter-structure distance connected with LIPSS. Along with the insight onto the physical picture driving the morphology evolution, our experiments also demonstrate that the resulting quasi-regular and random spiky morphology can be tailored by the intensity/polarization distribution of the incident laser beam allowing to achieve diverse hierarchical surface morphologies with reduced reflectivity in the visible and shortwave IR (SWIR) spectral ranges. In contrast to the well-studied fs-laser texturing of c-Si in air demonstrating similar combination of the electromagnetic and hydrodynamic processes driving morphology evolution, the methanol strongly reduces efficiency the silicon oxidation as well as allows to achieve larger density of nanostructures per surface cite defined by initially subwavelength LIPSSs periodicity. Finally, using the developed single-step fabrication strategy we produced a vertical p-n junction Si photodetector with an active laser-patterned area providing enhanced near-IR photoresponse and expanded operation window without annealing post-treatment under room temperature conditions and zero bias voltage.

\section{Results and discussion}

%________________________________fig 1
 \begin{figure*}
\center{\includegraphics[width=0.95\linewidth]{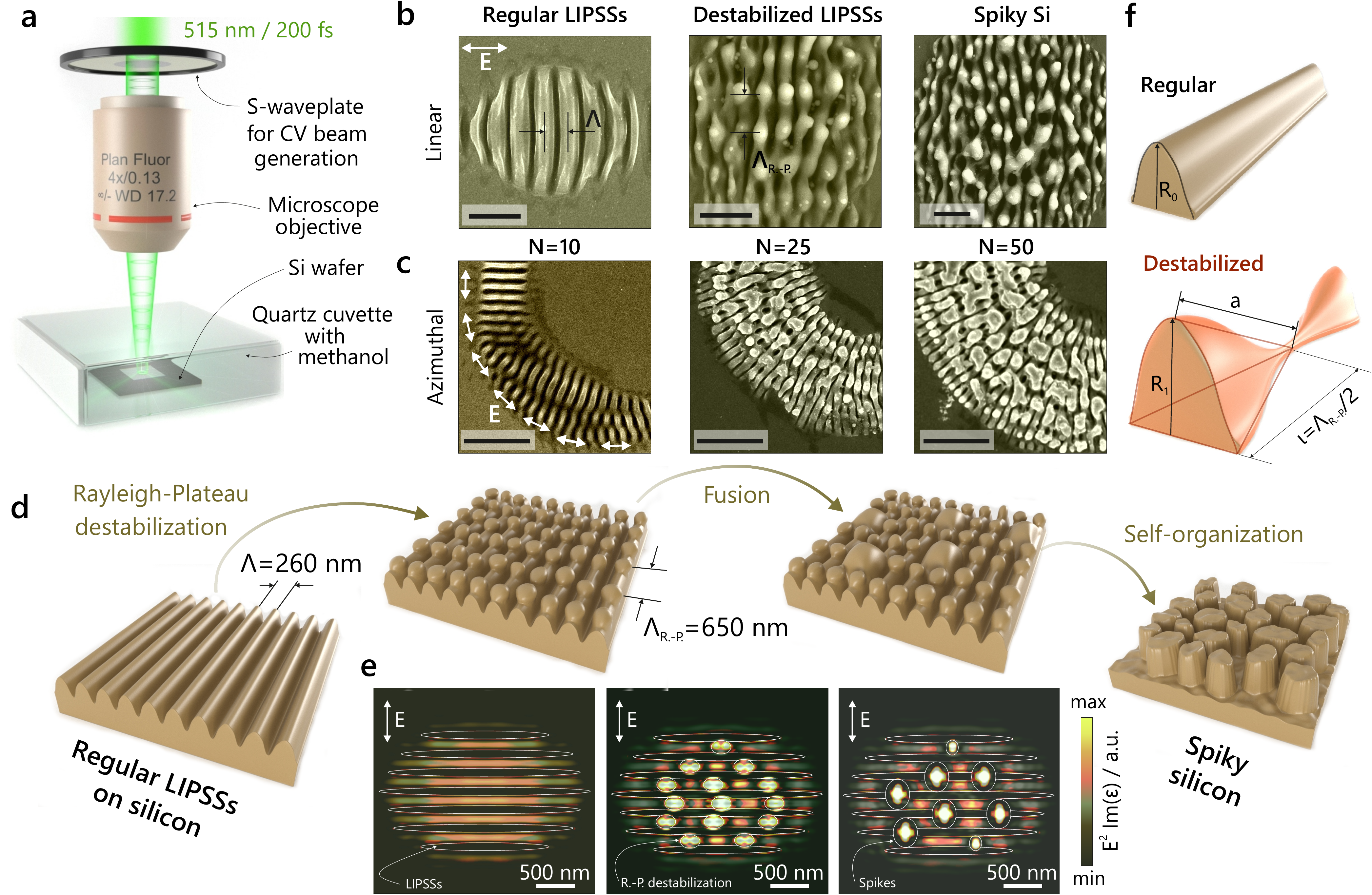}}
\caption{\space (a) Schematically illustrated procedure of fs-laser nanotexturing of monocrystalline Si in methanol. (b,c) Series of top-view SEM images of the spotted Si surface modifications produced by linearly polarized Gaussian beam and azymuthally polarized CV beam at fixed fluence of $F$=0.18 J/cm$^2$ and variable number of applied pulses per site $N$ (that is indicated near each image). Scale bar in panels (b) and (c) corresponds to 0.5 and 2 $\mu$m, respectively. One quarter of the circle-shape morphology produced with the CV beam is shown for clarity. (d) Sketch demonstrating key steps behind laser-driven morphology transition of the Si surface from the ordered subwavelength LIPSSs to self-organized spikes. (e) In-plane maps of the absorbed laser energy $\propto$E$^2\cdot$Im[$\epsilon$] in the vicinity of the LIPSS ridges of variable morphologies. White contour highlights the modeled morphologies (elongated elliptical pitches in the Si wafer and circularly shaped protrusions). Double-side arrows indicate the polarization direction. The obtained maps were averaged over several cross-sections calculated at variable depth below Si surface. (f) Schematic representation of one half of the period of the instability used for the estimations.}
\label{fig1}
\end{figure*}

\subsection{Formation of spiky Si under static irradiation conditions: onset of hydrodynamic processes in LIPSSs}

We started from systematic studies of the fs-laser ablation of Si in methanol under static (or spotted) irradiation conditions, where laser beams with either a Gaussian- or a donut-shape intensity profiles and variable polarization distributions (further referred to as Gaussian and cylindrical vector (CV) beams) were applied to expose certain sites of the Si surface (Fig. 1a). For simplicity, to illustrate the key processes underlying the morphology modulation, namely, transition from the LIPSSs to randomly arranged spiky morphology, we fixed laser fluence at $F$=0.18 J/cm$^2$ (that is larger than the single-pulse ablation threshold of Si in liquid; \cite{borodaenko2021deep}) and varied the number of applied pulses per site $N$. Under such irradiation conditions, any substantial modifications of the surface start at $N>$5 appearing as the subwavelength LIPSSs that exhibit periodicity $\Lambda$=260$\pm$30 nm and are arranged perpendicularly to the polarization direction (left panel, Fig. 1b). Consequently, azimuthally polarized CV beam imprints donut-shaped areas containing radially oriented LIPSSs (left panel, Fig. 1c). The LIPSSs were previously suggested to originate from interference of the incident radiation with surface plasma waves supported at the interface between surrounding liquid (methanol, in this case) and photoexcited layer of Si with the dielectric permittivity modified by increased concentration of free carriers \cite{borodaenko2021deep}. Subsequently incident laser pulses are absorbed by the LIPSS ridges resulting in their heating and partial melting. Molten ridges of a certain size are expected to become hydrodynamically unstable that results in modulations of their height with a characteristic period $\Lambda_{R.-P.}\approx$ 550-700 nm associated with the development of the R.-P. instability \cite{plateau1873statique,rayleigh1892xvi,kulchin2014formation,pavlov2020nanocrowns}. These height/width modulations are clearly seen on the surface morphologies imprinted with both Gaussian and CV beams at $N>$25 (middle panels, Fig. 1b,c) and are expected to facilitate light absorption of the subsequently incident laser pulses. This results in local fusion of the neighboring protrusions to spiky features. The process continues upon subsequent surface irradiation eventually leading to the formation of self-organized nanotextured surface covered with a randomly arranged protrusions (further referred to as spiky Si). Sketch provided in the Figure 1d graphically illustrates the main steps behind transition from the ordered subwavelength LIPSSs to self-organized textures. To justify the deductions regarding such laser-induced morphology evolution, its effect on the modification of the laser radiation absorption was first analyzed by numerically solving Maxwell equations with a commercial electromagnetic solver (Lumerical, Ansys). As initial step, we considered smooth Si LIPSSs in the form of the nanotrenches with semi-elliptical cross-section arranged with a periodicity of 260 nm perpendicularly to the polarization direction of the normally incident plane wave and calculated in-plane near-surface distribution of the absorbed laser energy $\propto$E$^2\cdot$Im[$\epsilon$] (E is an EM-field amplitude; $\epsilon$ is dielectric permittivity of silicon) in their vicinity. These calculations clearly show preferential absorption of the deposited laser energy by the LIPSSs ridges (left panel, Fig. 1e) as well as rearrangement of the absorbed energy profile once the R.-P. modulations and larger fused nanostructures appear (middle and right panels, Fig. 1e). In additional, both types of structures appeared to enhance laser radiation absorption promoting material melting and fusion.

Let us next justify the onset of the R.-P. instability developing in the molten LIPSS ridges with a characteristic period $\Lambda_{R.-P.}\approx$ 550-700 nm according to the experimental data (middle panels; Fig. 1b,c). We considered a ridge having at the initial state (i.e., before the instability starts developing) a half-cylindrical shape, as shown in the top panel of Fig. 1f, while at the final state each half-period of the destabilised ridge was approximated with a half-cone. The precise estimation of the Rayleigh-Plateau destabilisation requires the information about the contact angle of liquid silicon on solid silicon substrate\cite{pavlov2020nanocrowns}, which is not known and most probably cannot be precisely identified due to the presence of silicon dioxide \cite{SiCA}. Hence, we assessed the characteristic destabilisation time $t_i$ and possible range of the characteristic periods $\Lambda_{R.-P.}$ based on the comparison between the surface energy in the initial and final states. In doing this, we supposed that the volume of the considered ridge does not change during the R.-P. destabilisation enabling the estimation of the ratio between the initial radius $R_0$ and the maximal radius of the cone in the final state $R_1$. From $\frac{1}{2}\pi R_0^2\ell = \frac{1}{2}\frac{1}{3}\pi R_1^2\ell$ follows that $R_1=\sqrt{3}R_0$. A destabilisation is only possible if it is energetically advantageous for the system. Only such periods of the instability $\Lambda$ can develop, for which the surface energy decreases upon the shape transformation. No other kinds of energy are considered, since the surface tension dominates on the submicrometer scale. The surface energy of the side surface of a half-cylinder is $\mathcal{E}_0=\pi R_0\ell\sigma$ and in the final state the energy of the side surface of the half-cone is $\mathcal{E}_1=\frac{1}{2}\pi R_1a\sigma$, where the surface tension of liquid silicon $\sigma\approx0.7\,...\,0.8\,N/m$ \cite{EUSTATHOPOULOS201377} and the side wall of the cone is $a=\sqrt{\ell^2+R_1^2}$. From the conditions that the energy difference must be $\mathcal{E}_0-\mathcal{E}_1>0$ and $\ell=\Lambda_{R.-P.}/2$ follows that only periods with $\Lambda_{R.-P.}>6R_0$ are possible. This estimation well fits the experimentally observed periods of the instability $\approx 0.6\,...\,0.7\,\mu m$ at $R_0\approx 0.1\,\mu m$.

 For the estimation of time $t_i$ the instability with a given period $\Lambda_{R.-P.}$ is needed to develop, we supposed that the gain in the energy due to the reduction of the surface limits the work of the friction forces $A=Fs$, which appears due to the motion of the viscous melt \cite{gurevich2016mechanisms}. Here $s\sim\ell$ is the distance, at which Si atoms are transported upon the instability, and the friction force is estimated with the Newton's law as $F=\eta A v/R_0$, where the viscosity of liquid silicon $\eta=0.7\,...\,0.9\,mPa\,s$ \cite{Sasaki1995viscosity}, the average flow velocity of the melt on the surface is $v\approx s/t_i$, and the area of the moving liquid $A\approx \pi R_0\ell$. With $\ell=\Lambda_{R.-P.}/2$ this criterion can be rewritten as:
\begin{equation}\label{Eq:Time1}
 \dfrac{\eta\pi\Lambda_{R.-P.}^3}{8t_i} \lesssim \sigma\pi R_0\dfrac{\Lambda_{R.-P.}}{2}\left(1-\dfrac{\sqrt{3}}{2}\left(1+12\dfrac{R_0^2}{\Lambda_{R.-P.}^2} \right)\right).
\end{equation}

The term in the brackets in the equation~\eqref{Eq:Time1} is of the order of one, and the equation can be simplified as
\begin{equation}\label{Eq:Time2}
 t_i\gtrsim \dfrac{\eta\Lambda_{R.-P.}^2}{\sigma R_0}.
\end{equation}

Substituting the experimentally observed period of $\Lambda_{R.-P.}\approx 0.6\,...\,0.7\,\mu m$ we obtain that the instability develops within few nanoseconds that agrees with the experimental assessments of the characteristic solidification time $t_m$ during which the Si exposed with a fs-laser pulse remains in the molten state, thus allowing the instability to develop \cite{Gundrum2007Sisolodification}. This time can be roughly estimated based on the assumption that the temperature in the molten Si ridge is $T_1\approx 2T_m$, while the temperature of the solid substrate in the proximity of the molten ridge $T_2\approx T_m$, where the $T_m$ is the Si melting point. We supposed that after the end of the laser pulse the heat conduction is the only mechanism of cooling and use the Fick's laws of diffusion, in which the Laplace operator is approximated as $\nabla^2T\approx T_m/R_0^2$ and the time derivative of the temperature $\dot{T}\approx T_m/t_m$. With the thermal diffusivity in liquid silicon $D_T\approx 0.1\,cm^2/s$ \cite{Yamamoto1991} we estimate $t_m\approx R_0^2/D_T\sim 10^{-9}\,s$, which is of the order of estimated  time required for the development of the instability and agrees with the previously reported data  \cite{Gundrum2007Sisolodification}.

According to the performed analysis, the development of the instabilities with larger periods is evidently limited by the solidification time $t_m$  as well as the characteristic LIPSSs morphology (periodicity and ridge size; Eq. ~\eqref{Eq:Time2})). This means that the regulation of the laser-processing conditions that define the surface morphology at initial state (i.e. before the instabilities start to develop) allows to control the resulting morphology of the self-organized textures, such as average size and density of the spiky features. At the same time, larger applied fluences as well as destabilization of the LIPSSs through a fusion-assisted formation of the higher light-absorbing features evidently will increase the size of the molten pool and its solidification time making possible the development of the instabilities with larger wavelengths (Fig. 1e).

\subsection{Formation of spiky Si under dynamic irradiation conditions: correlation between irradiation conditions, nanomorphology evolution and optical properties}

%________________________________fig 2
 \begin{figure}
\center{\includegraphics[width=0.98\linewidth]{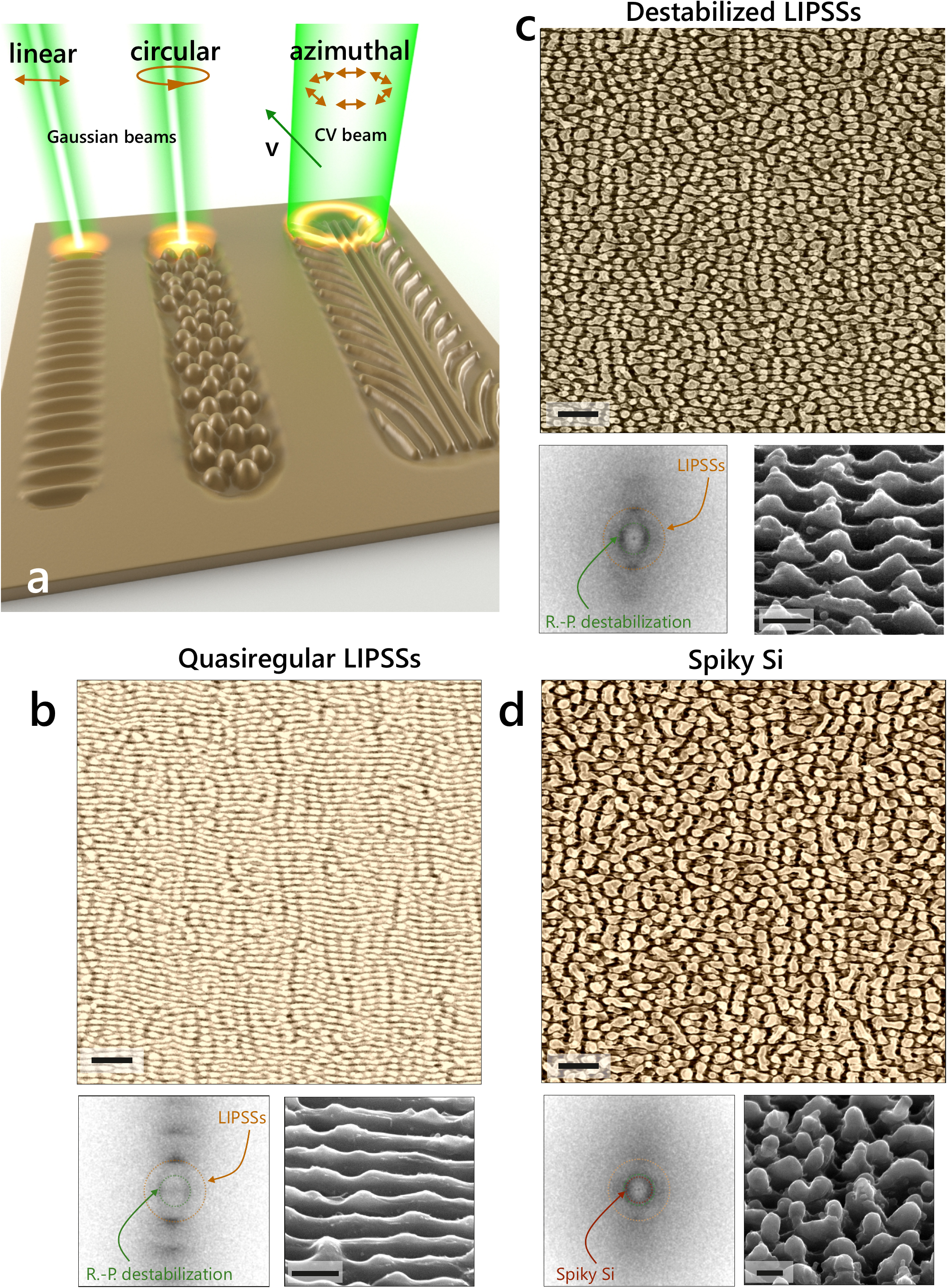}}
\caption{\space (a) Schematic illustration of the Si wafer texturing in methanol using dynamic irradiation conditions and various laser beams. (b-d) Evolution of the surface morphology of the Si wafer processed in methanol by Gaussian-shape linearly polarized fs-laser pulses at $F$=0.146 J/cm$^{2}$ (regular LIPSSs), 0.155 J/cm$^{2}$ (destabilized LIPSSs) and 0.177 J/cm$^{2}$ (spiky Si). Bottom panels provide FFT analysis of the main image as well as close-up tilt-view SEM images of the corresponding surface morphologies. Fixed scanning speed $\nu$=50 $\mu$m/s and pulse repetition rate $\kappa$=1 kHz were used for surface patterning. Scale bar in top-view and tilt-view SEM images in the panels (b)-(d) corresponds to 2 and 0.5 $\mu$m, respectively.}
\label{fig2}
\end{figure}

The observed surface morphologies including regular and destabilized LIPSSs as well as disordered spiky Si can be expanded over practically relevant areas under dynamic irradiation conditions, namely, by scanning the Si surface by a laser beam diameter $d$ at a constant speed $\nu$ and pulse repetition rate $\kappa$ (Fig. 2a). In fact, both parameters define the number of pulses $N$=$d\kappa$/$\nu$ applied per surface site, similarly to the case of static irradiation conditions discussed in the previous section. Moreover, once the instability development is related to the amount of the molten phase of the material, tailoring the laser fluence $F$ within a certain range above the ablation threshold also allows to tune the resulting surface morphology. Series of top- and side-view SEM images in Fig. 2b-d illustrate the ability to reproduce the main discussed morphologies by scanning the Si surface with a linearly polarized Gaussian beam along meander-like trajectory at the inter-line spacing of 2.5 $\mu$m, $\kappa$=1 kHz, $\nu$=50 $\mu$m/s and varying $F$ (Fig. 2b-d). The obtained SEM images of the surface morphologies were also subjected to Fast Fourier Transform (FFT) analysis clearly showing peaks related to the regular LIPSSs as well as those associated with the periodic modulations caused by the R.-P. instability (Fig. 2c,d). FFT analysis shows that the intensity of the latter peaks grows with the applied fluence $F$, while protrusions mainly follow the direction perpendicular to the LIPSS orientation. At laser fluence of $F$=0.18 J/cm$^2$ completely disordered spiky Si surface can be produced as evidenced from the FFT analysis (Fig. 2d).

Performed analysis also highlights that upon increasing the number of applied pulses electromagnetic processes underlying formation of the ordered LIPSSs are gradually replaced by the hydrodynamic ones that evidently dominate at larger $N$. Noteworthy, further increase of the number of applied pulses (and/or laser fluence $F$) promotes melting and fusion of the neighboring protrusions resulting in increase of their geometric size and corresponding decrease of their density. In fact, increase of the laser fluence $F$ from 0.166 (minimal fluence required to achieve spiky Si at $\nu$=50 $\mu$m/s) to 0.199 J/cm$^2$ decreases twice the density of the spiky features (from 2.9 to 1.53 structures per $\mu$m$^2$).
%\red{\texttt{Future plans: Can we check how the average distance between these structures grows with the time (number of pulses)? The theory predicts $L\propto t^{\beta}$ with $\beta\lesssim2/5$ if the coarsening happens along one ridge and $\beta\lesssim3/8$ if the coarsening happens on a plane substrate [Felix Otto, Tobias Rump, and Dejan Slepcev, "Coarsening Rates for a Droplet Model: Rigorous Upper Bounds", SIAM Journal on Mathematical Analysis, Vol. 38, Iss. 2 (2006) 10.1137/050630192]}}
The increase in the average distance between the spiky features with time agrees with the theory of coarsening droplets on a plane predicting the distance growth $\propto t^{\beta}$ with $\beta\lesssim3/8$ \cite{OttoDrops}. At the same time, at moderate fluence (or $N$) the simultaneous contribution of electromagnetic and hydrodynamic processes drives formation of unique quasi-regular hierarchical morphologies, where LIPSSs ridges act as sites for templated dewetting \emph{via} development of the R.-P. instability resulting in the largest density of structures per surface area.

%________________________________fig 3
 \begin{figure}
\center{\includegraphics[width=0.98\linewidth]{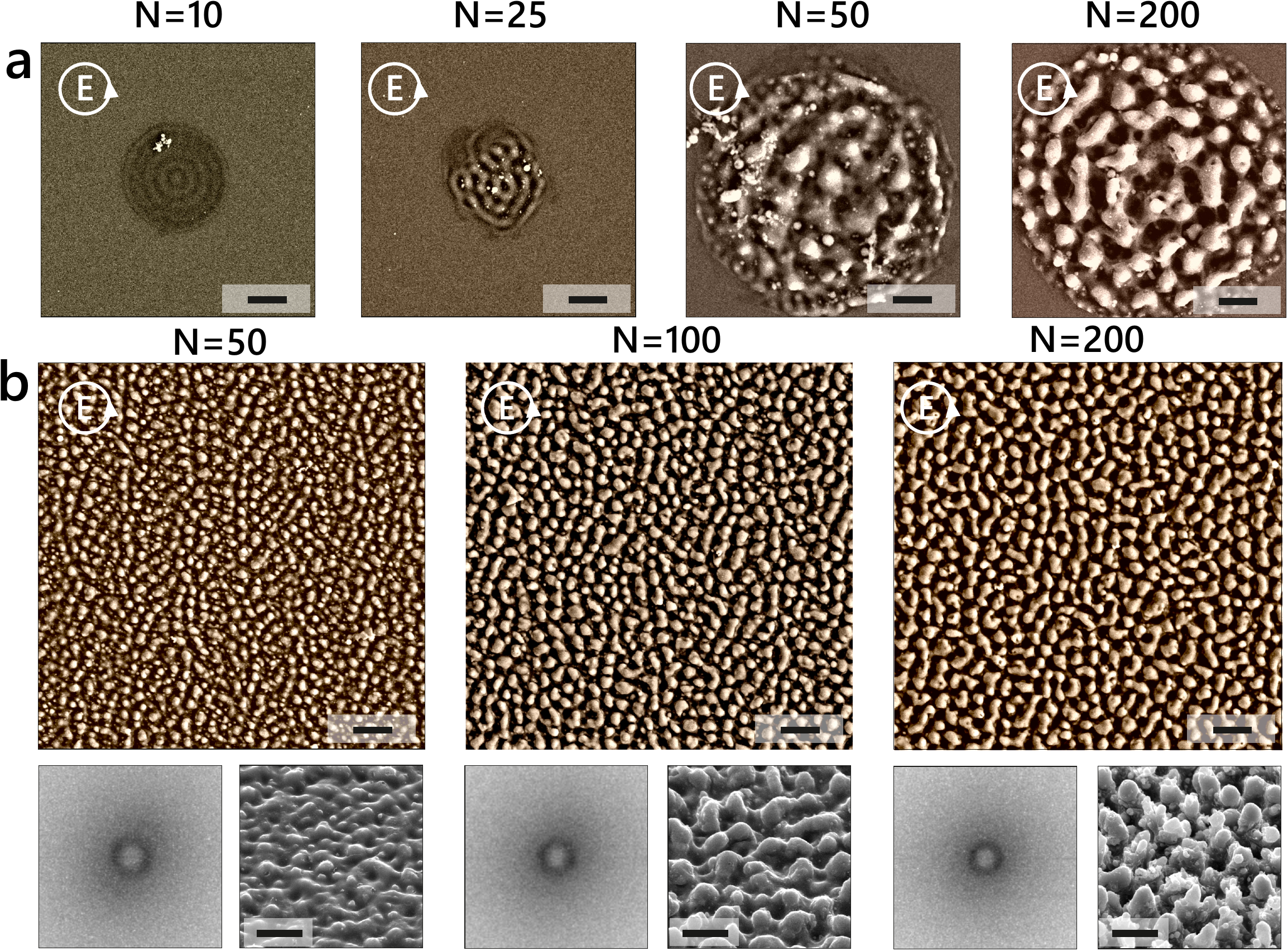}}
\caption{\space (a) Series of top-view SEM images of the spotted Si surface modifications produced by circularly polarized Gaussian beam at fixed fluence and variable number of applied pulses per site $N$. (b) SEM images of the Si surface morphology processed in methanol by circularly polarized fs-laser pulses at variable number of applied pulses per site $N$. Bottom panels provide FFT analysis of the main image as well as close-up tilt-view SEM images of the corresponding surface morphologies. Fixed laser fluence $F$=0.16 J/cm$^2$ and pulse repetition rate $\kappa$= 1 kHz was used for surface patterning. Scale bars in panels (a) and (b) corresponds to 1 and 2 $\mu$m, respectively.}
\label{fig3}
\end{figure}

Once the LIPSSs formation governs the subsequent morphology evolution, it is reasonable to suppose that control over the morphology at the initial step (that is largely defined by electromagnetically driven self-organization processes) will substantially affect the resulting relief at the steps when hydrodynamic processes will come into action. To support this deduction, we carried out parametric studies on the Si wafer patterning in methanol using circularly polarized Gaussian-shaped beam under static and dynamic irradiation condition. Such irradiation conditions suppress the efficiency of the LIPSSs formation along a certain direction as well as deepening of the nanotrenches by subsequently incident pulses. This leads to the formation of the shallow LIPSSs with azimuthally varying orientation at $N$=10 and $F$=0.16 J/cm$^2$ (Fig. 3a, left panel). Noteworthy, similar relief was obtained by double-pulse exposure of the Si wafer using ps-scale time delay between pulses and orthogonal polarization direction \cite{liu2021formation}. The produced morphology is much less pronounced being compared to those obtained by linearly polarized beam and same irradiation conditions ($N$ and $F$) (see Fig. 1b). Subsequent laser exposure of this morphology promotes formation of the disordered LIPSSs which height modulation can be driven by the interference of the diverging/converging plasmon waves or R.-P. instabilities with $\Lambda_{R.-P.}\approx\Lambda$ defined by reduced size of the ridges. Eventually, further irradiation ($N>$50) initiates merging of the neighboring elevations resulting in formation of rather smooth surface protrusions, which size increases $via$ fusion upon subsequent exposure. Similar surface morphologies were also produced with circularly polarized Gaussian beam under dynamic irradiation conditions at fixed $F$ and $\kappa$ varying scanning speed $\nu$ to provide the number of applied pulses $N$ per site ranging from 50 to 200 (Fig. 3b). These morphologies were analyzed with the FFT revealing their complete directional independence as well as constant average distance between spiky features that raises with $N$. In contrast to the surface patterning with the linearly polarized beam, utilization of the circularly polarized one can not provide regimes where evident formation of the LIPSSs destabilized by the R.-P. instability can be justified. Additionally, the obtained morphologies exhibit more shallow and smooth profile under irradiation conditions similar to those used with linearly polarized beam, revealing clear difference in the onset of hydrodynamic processes affected by the LIPSSs formation at initial steps that confirms the previously made deductions.

%________________________________fig 4
 \begin{figure}
\center{\includegraphics[width=0.9\linewidth]{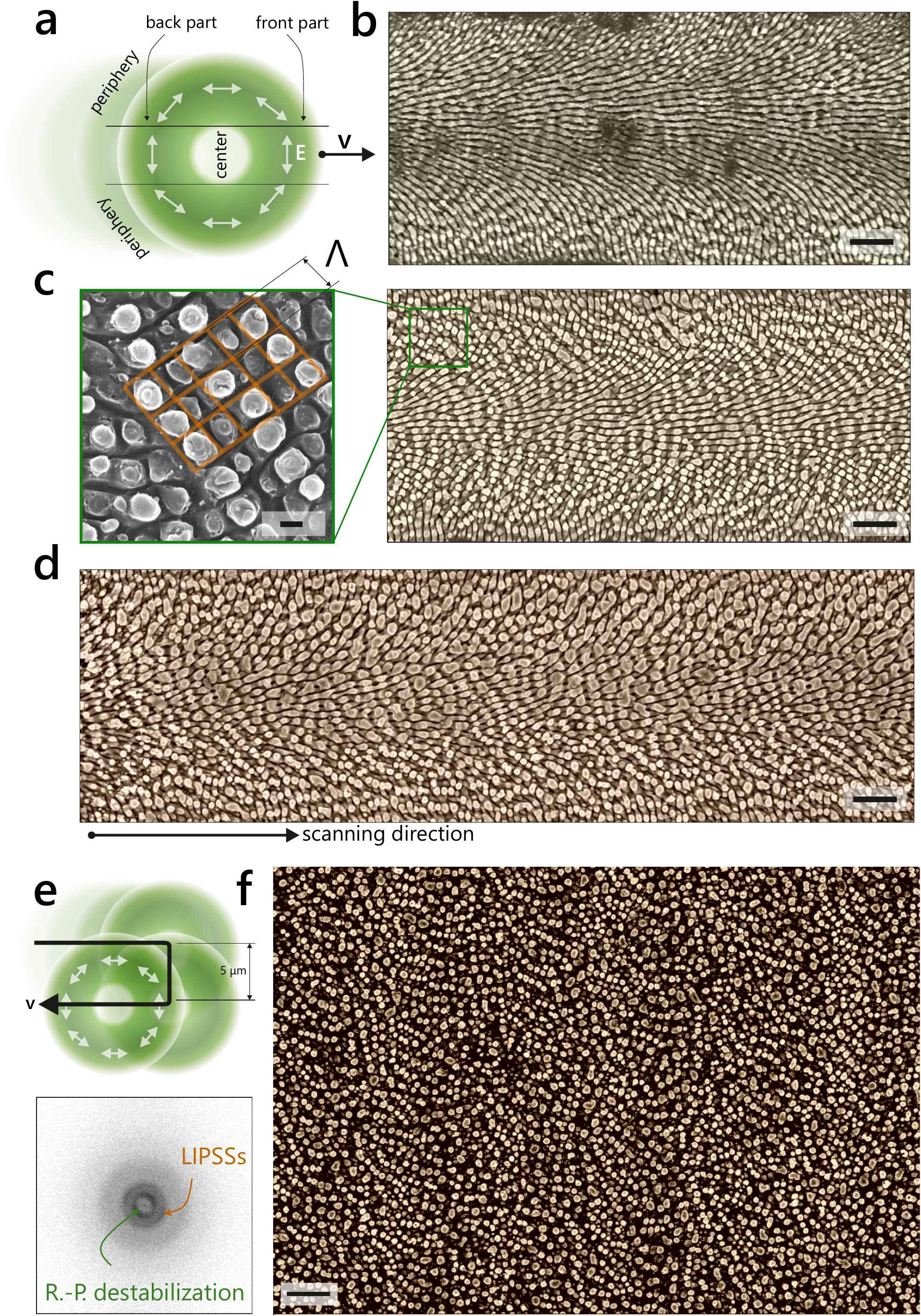}}
\caption{\space (a) Schematic illustration of the CV beam clarifying the origin of the inhomogeneous patterning of the Si surface under dynamic irradiation conditions (scanning along a linear trajectory). (b-d) SEM images of the Si surface patterned in methanol with an azimuthally polarized CV beam at fixed fluence $F$=0.16 J/cm$^2$ and variable scanning speed $\nu$= 150, 100, 50 $\mu$m/s respectively. (e) Scheme of the uniform Si surface patterning with a CV beam at an offset between the linear scans equal to the half of the beam diameter (approx. 5 $\mu$m). (f) SEM image of the surface morphology produced using such scanning procedure, $F$=0.16 J/cm$^2$ and $\nu$=100 $\mu$m/s as well as FFT analysis of this image. Scale bar corresponds to 2 $\mu$m on the main panels and 0.2 $\mu$m on the inset.}
\label{fig4}
\end{figure}

Similar Si patterning experiments were also carried out with an azimuthally polarized CV beam under dynamic irradiation conditions revealing even broader diversity of the obtained surface morphologies (see Fig. 4). In particular, near-threshold fluence and moderate number of applied pulses allows to produce the patterned tracks containing the radially arranged LIPSSs. Once the $F$ (or $N$) increases, the observed grating-type morphology of the peripheral areas of the track changes to densely arranged spiky surface. The feature can be explained by the donut-shape intensity profile of the CV beam, allowing to sequentially expose the Si surface by the front and back parts of the donut upon scanning along linear direction. In the central parts of the linear scans the polarization distribution appears to be identical for the front and the back parts of the donut, being almost perpendicular to the scanning direction. However, for peripheral area the front and the back part of the donut have different polarization direction as it is schematically illustrated in Fig. 4a. Evidently, in the peripheral areas of the scan this creates two overlapped LIPSS gratings with almost perpendicular orientations resulting in spiky surface with the largest density of the surface features per surface area (Fig. 4b). Further increase of the dose ($F$, $N$) leads to the previously discussed hydrodynamic R.-P. destabilization and fusion-assisted formation of larger surface features. Noteworthy, unveiled nanotexturing regime with the overlapping LIPSSs can be expanded over larger surface areas upon scanning the surface with CV beam along the meander-like trajectory and keeping the offset between the linear scans equal to the half of the CV beam diameter (approx. 5 $\mu$m; Fig. 4e). The resulting surface morphology produced at $F$=0.16 J/cm$^2$ and $\nu$=100 $\mu$m/s as well as FFT analysis of this image are illustrated in Fig. 4f indicating presence of both the regular and hydrodynamically destabilized LIPSSs with a random orientation in the formed surface pattern.

%________________________________fig 5
 \begin{figure*}
\center{\includegraphics[width=0.85\linewidth]{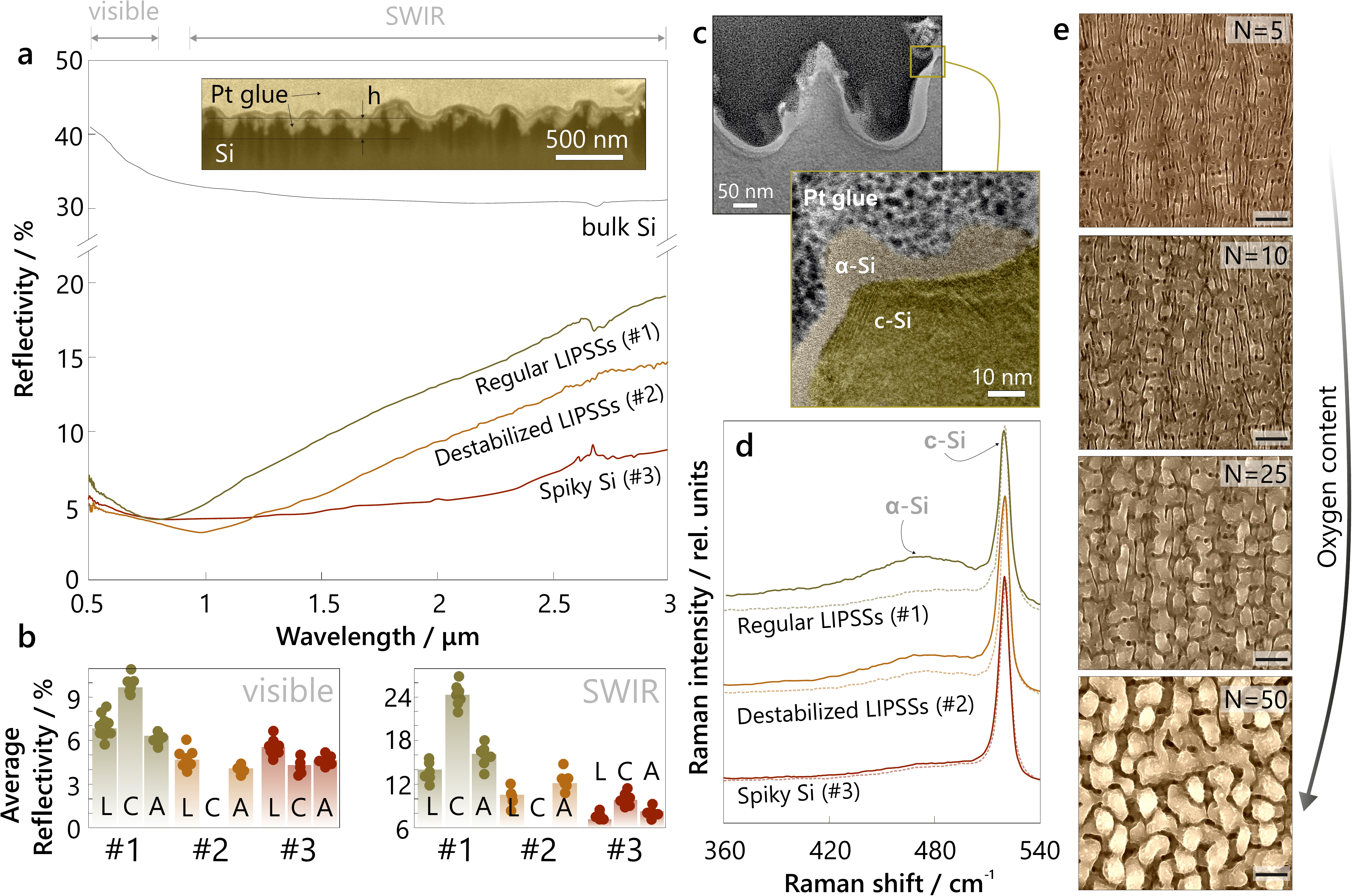}}
\caption{\space (a) Representative FTIR reflectance spectra of the pristine and laser-patterned monocrystalline Si. Various types of the surface morphologies (regular \#1 and destabilized LIPSSs \#2 as well as the spiky Si \#3) were fabricated at $F$=0.16 J/cm$^2$ and $\kappa$=1 kHz varying only the scanning speed $\nu$ controlling the number of applied pulses per site $N$. Inset shows SEM image of the FIB cross-sectional cut made perpendicularly to the ridge orientation. (b) Average visible and SWIR reflectivity of various types of the surface morphologies (regular \#1 and destabilized LIPSSs \#2 as well as the spiky Si \#3) produced using linearly (L), circularly (C) and azimuthally (A) polarized beams and variable $F$ and $\nu$. (c) High-resolution TEM image of the LIPSS ridges with their well-seen amorphous interface layer.(d) Corresponding averaged Raman spectra of the laser-patterned Si surface sites related to the mentioned morphologies before (solid curves) and after (dashed curves) annealing at 600$^o$C and 10$^{-5}$ torr for 30 min. (e) Series of SEM images of the Si surface processed in air at variable number of applied pulses $N$ and fixed $F$=0.16 J/cm$^2$. Scale bar corresponds to 2 $\mu$m.}
\label{fig5}
\end{figure*}

The produced spatially homogeneous surface morphologies were further analysed through the way they interact with the incident unpolarized electromagnetic waves. More specifically, we first compared the average reflectivity of the laser-patterned surfaces produced with a linearly polarized beam in the visible (0.5 - 0.7 $\mu$m) and shortwave IR (SWIR; 1 - 3 $\mu$m) spectral regions (Fig. 5a). These studies showed that SWIR reflectivity gradually drops once the morphology evolves from the regular subwavelength LIPSSs to the spiky Si, which reflects the formation of the larger supra-wavelength structures. Noteworthy, in the SWIR spectral range the regular LIPSSs satisfy the criterion of the subwavelength grating, for which at wavelengths $\lambda_R$ larger than $\Lambda n_{Si}$ ($n_{Si}\approx$3.5 is refractive index of Si in the SWIR spectral range \cite{aspnes1983dielectric}) this surface morphology can be approximated by an effective medium with a gradually decreasing refractive index when moving from the air interface towards the bulk. In this respect, the removal of the refractive index jump decreases the Fresnel reflection at the interface. However, optimal anti-reflective performance can be achieved when the thickness of this effective medium defined by the height of the surface structures $h$ will be larger than $\lambda_R$(4$\sqrt{n_{Si}})^{-1}$. Considering the average height of the ridges $\approx$ 200 nm for regular LIPSSs (inset; Fig. 5a), this explains the increase of the reflectivity at $\lambda_R>$1 $\mu$m. Onset of hydrodynamic processes allows to create dense supra-wavelength surface features that efficiently traps the IR light through multiple reflections resulting in average reflectivity R$_{SWIR}\approx$ 6\% for spiky Si ($vs$ only $\approx$11 and 14 \% for destabilized and regular LIPSSs, respectively; Fig. 5b). Noteworthy, all the considered types of the structures demonstrate rather low reflectivity in the visible spectral range (down to 5\% for destabilized LIPSS; Fig. 5b), that can be associated with their light trapping performance as well as certain structural transformation of the near surface layer confirmed by combining transmission electron microscopy and Raman microspectroscopy (Fig. 5c,d). In particular, both methods revealed the formation of thin amorphous Si layer ($\alpha$-Si) with its thickness being about 10-25 nm for the regular LIPSSs according to TEM studies. Moreover, systematic Raman studies showed that the $\alpha$-Si amount decreases with the increasing amount of laser pulses applied per surface site (decreasing $\nu$) as evidenced from gradually decaying intensity of the corresponding Raman band centered near 480 cm$^{-1}$ (Fig. 5d; solid curves) \cite{borodaenko2021deep,liu2023one}. Importantly, this observation indicates that promoted laser exposure creates continuously growing amount of the molten Si that recrystallizes over longer time period required for minimization of the amorphous phase. In this respect, spiky Si morphology exhibits the smallest amount of the $\alpha$-Si almost unaffected by annealing of the patterned Si wafers in an oxygen-free (10$^{-5}$ torr) atmosphere at 600$^o$C for 30 min (dashed curves; Fig. 5d). As can be seen, effect of the annealing is more evident for regular and destabilized LIPSSs where initial content of the $\alpha$-Si is larger. Systematically studied SWIR and visible-range reflectivity of the main types of the surface structures (marked as \#1 - \#3 for regular and destabilized LIPSSs as well as the spiky Si) produced using various laser beams with linear (L), circular (C) and azimuthal (A) polarization within a certain range of laser processing parameters ($F$,$\nu$) are summarized in Fig. 5b revealing the ability to control optical properties of the Si surface within the demanding spectral range upon proper surface micro- (nano-)texturing.

Finally, let us highlight additional benefits of the Si laser texturing in methanol. In fact, most of the studies on fs-laser processing of Si was carried out under ambient environment revealing ability to create supra-wavelength structures (also referred to as grooves) under accumulated exposure. These structure manifested themselves as a universal phenomenon being observed under diverse experimental conditions (such as visible and IR laser wavelengths \cite{allahyari2020formation}, angled excitation \cite{nivas2021incident}, shaped beams \cite{nivas2015laser}, GHz bursts \cite{kawabata2023two}, vacuum conditions \cite{hu2022ultrafast}, etc.) with their supra-wavelength periodicity generally correlating with the incident wavelength \cite{nivas2018direct}. These structures at the initial step of their formation are believed to originate from the similar hybrid electromagnetic/hydrodynamic scenario unveiled here, yet alternative explanation based on hydrothermal waves were also reported \cite{tsibidis2015ripple}. Irrespectively of the chosen laser wavelength and processing parameters, the reported characteristic distance between the supra-wavelength grooves that evidently defines the average density of the spiky features in between is about twice larger compared to the corresponding LIPSS period $\Lambda\approx$0.8$\lambda$ upon processing in air. Taking into account the observed LIPSS periodicity of $\Lambda\approx$0.55$\lambda$ and suggested scenario of the morphology evolution $via$ destabilization and fusion of the neighboring LIPSS ridges, more denser arrangement of the spiky features are expected for Si surface processed in methanol. To provide direct comparison, we performed laser texturing experiments of Si wafer using similar processing parameters under ambient environment and static/dynamic irradiation conditions (Fig. 5e). These results confirm the made deductions regarding (i) similarity of the physical processes driving morphology reorganization revealed under dynamic exposure that results in (ii) notably lower maximal density of the spiky features (less than 1 structure per $\mu$m$^2$; Fig. 5e) in the textures produced in air. Moreover, increase of the dose results in contentiously growing amount of the oxidized Si that can be crucial for certain applications and is to be removed by ultrasonication or HF etching, in a sharp contrast to the surface textures produced in methanol requiring no additional cleaning procedures. Comparative energy-dispersive X-ray (EDX) analysis indicates an order of magnitude larger content of oxygen (30$\pm$7 wt.\% $vs.$ 3  wt.\% at $N$=50, $F$=0.16 J/cm$^2$) in the surface textures patterned under ambient environment, with a content of oxygen in the pristine Si wafer stored in air of 2 wt.\%. Additionally, processing in methanol leads to an small increase of the carbon content only to 6 wt.\% ($vs.$ 5 wt.\% for non-patterned Si wafer) that can be related to a thermal decomposition of the methanol molecules or substitution of the OH-groups on the silicon surface with -OCH$_3$ \cite{SiOCH3}. %All these points clearly highlight benefits of utilizing methanol as an environment for Si nanopatterning towards diverse applications in sensors, optoelectronic and photo-detection.

\subsection{Laser-patterned Si p-n junction photo-detector}

Discussed benefits of the liquid-assisted laser texturing of Si highlight practical attractiveness of the developed approach for diverse applications. For example, nanotextured Si surface can be further functionalized with noble-metal films (nanoparticles) to act as a substrate for optical sensing, where the ability to control morphological/optical properties are highly demanded to optimize the device performance for a certain pump wavelength \cite{diebold2009femtosecond,borodaenko2022demand,erkizan2022lipss}. Alternatively, metal-free sensing and light-emitting devices empowered by the rich optics of Mie resonances and benefiting from  the absence of luminescence quenching effects are also becoming popular  research direction \cite{alessandri2016enhanced,tittl2018imaging,mitsai2018chemically,mironenko2019ultratrace,dostovalov2020hierarchical}. Silicon also represent an important material applied in solar cells and photodetectors where the proper laser-assisted nanotexturing can not only remove optical coupling losses associated with an inherently high refractive index of the material, but also provide additional defect-mediated light-absorption channels boosting the device performance \cite{sher2011pulsed,zielke2012direct}.

%________________________________fig 6
 \begin{figure}
\center{\includegraphics[width=0.95\linewidth]{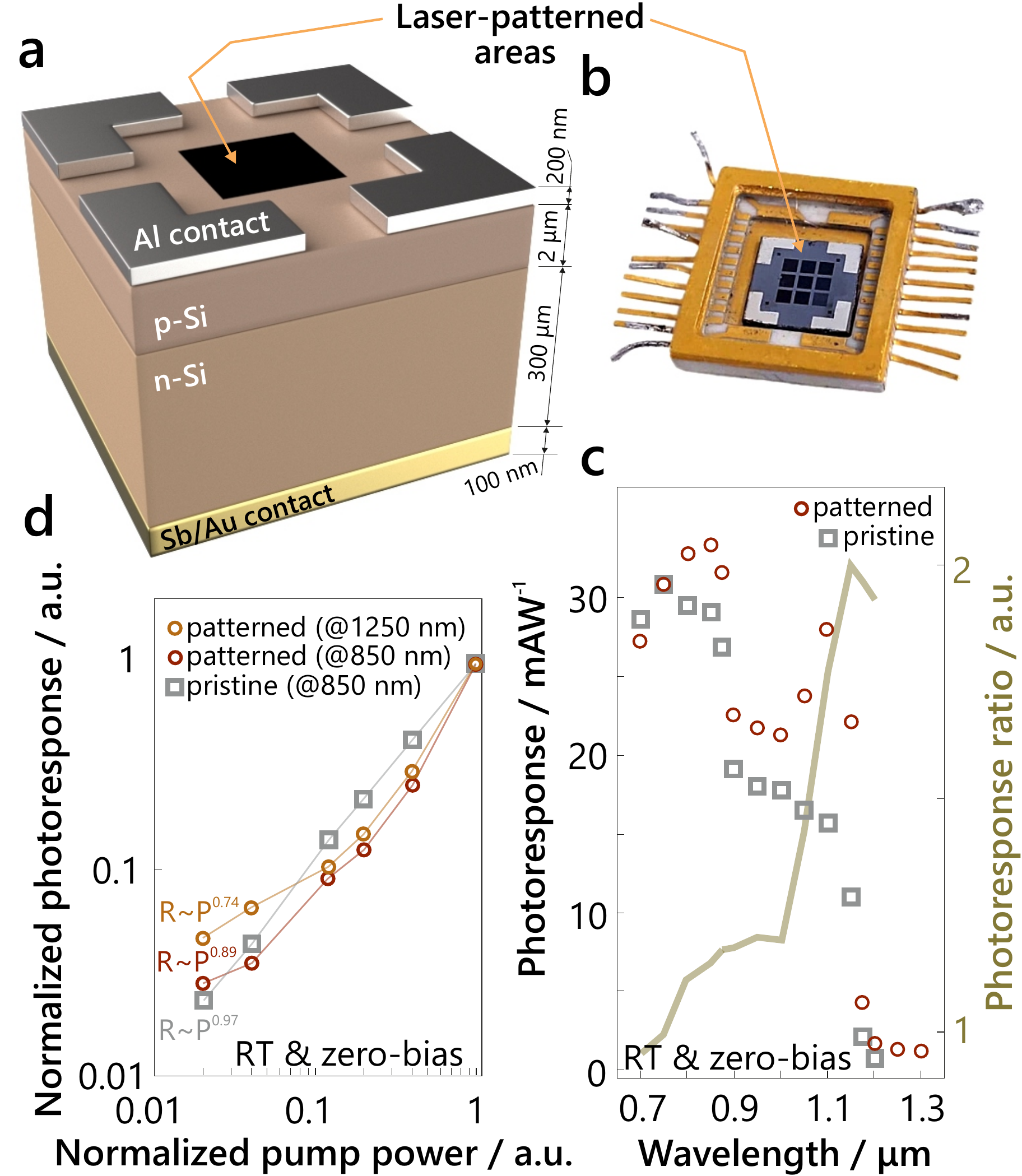}}
\caption{\space (a) Schematically illustrated vertical p-n junction Si photodetector with a laser-patterned active area. (b) Optical photograph of the fabricated device. (c) Spectral dependence of the photoresponse (markers) of the pristine $R_{flat}$ and laser-patterned $R_{spiky}$ devices. Solid curve demonstrates spectral dependence of the photoresponse ratio $r$=$R_{spiky}$/$R_{flat}$. (d) Normalized photoresponse as a function of the laser pump power measured for pristine device at 850 nm pump wavelength (gray markers) as well as for laser-patterned device at 850 (red markers) and 1250 nm pump wavelengths (orange markers).}
\label{fig4}
\end{figure}

We justified applicability of the developed liquid-assisted fs-laser texturing of Si for improvement of the characteristics of the p-n junction Si photodetector. Usually, extending the operation wavelength range and/or enhancement of the absolute photoresponse are the most demanded outputs, that are commonly achieved by nanostructuring-induced near-surface (shallow) junction formation \cite{juntunen2016near,garin2020black,seta?la?2023boron} and Si hyperdoping accompanied with surface texturing \cite{zhao2023black,zhao2020ultrafast} when dealing with low photo-detection performance of the pristine (unprocessed) Si-based devices in the UV and NIR/SWIR ranges, respectively. %Here, we used spiky Si type of nanostructuring for the processing of the Si p-n junction device, since resulted pattern effectively suppresses reflectivity from VIS to SWIR wavelength range, while associated strong incident light localization ("hot spots") promotes higher optical absorption thus enhancing the photo-current generation.
Scheme of the resulted Si photo-detector and photograph of the chip-packaged device with laser-patterned light-sensitive areas are presented in Fig. 6a,b. For device realization, we choose the destabilized LIPSS morphology providing the lowest reflectivity near 1000 nm as well as moderate morphology-affected depth that is to be carefully control considering low thickness (approx. 2 $\mu$m) of the top p-Si. Excessive texturing of this layer can deteriorate the crystallinity of p-n junction area (space charge region near the interface between n- and p-Si layers).

%The key feature of using manufactured p-n junction compared to hyperdoping or induced junction approach is in an easy and precise control on how deep will be resulted spiky morphology by varying the number of pulses. It allows to pattern only the top Si layer (p-Si) without any significant influence on the bottom Si (n-Si).
%In other words, this regime does not deteriorate , while enhances localization of light whereon, which is favorite for an effective separation of photo-generated carriers by p-n junction built-in electrical field.

Measurements of the photoresponse as a function of the incident laser wavelength (Fig. 6c) revealed the improved performance of the laser-textured device (R$_{spiky}$) over pristine one (R$_{flat}$). For clarity, the obtained data is also plotted as photo-response ratio, r=R$_{spiky}$/R$_{flat}$, showing its fast increase with the incident light wavelength, with the maximal 2-fold enhancement of the photoresponse at 1200 nm. Importantly, the laser patterning makes the Si p-n junction to be sensitive to sub-band gap illumination at least down to 1300 nm with the room-temperature photoresponse of about several mA/W under zero bias condition (self-powered mode). The mechanism of such improved NIR photo-sensitivity can be attributed to Urbach states and laser-induced structural defects generated in morphology-modified and underlying layers during the laser processing \cite{li2018sub}. This results in a lower photo-response of laser-patterned device as compared to the pristine one, when shifting to the shorter wavelengths due to the pronounced surface recombination rate for photo-generated carriers near the edge and side walls of the Si protrusions facilitated by lower penetration depth of the laser radiation. In accordance with Fig. 6c the specific threshold wavelength is 750 nm, where photoresponse ratio is equal to 1. As wavelength of the incident light exceeding threshold value becomes comparable with characteristic period of the Si featured ($\approx\Lambda_{R.-P.}$), the higher photoresponse could be detected. This period matching results in spectral shifting of the photoresponse peak value from the 750 to 850 nm with an additional local maximum at $\approx1100$ nm for the flat and spiky Si p-n junction photo-detectors, respectively. An additional confirmation of the defect-related enhancement of the NIR photoresponse is provided in Fig. 6d, demonstrating dependencies of the normalized photoresponse as a function of the incident laser power measured for both patterned and pristine devices at 850 and 1250 nm characteristic pump wavelengths. As can be seen, at photon energy above the Si bandgap (850 nm pump), both devices demonstrate power exponents close to 1, yet the data obtained for the laser-patterned device shows slightly sub-linear trend, evidently owing to enhanced surface recombination. At the same time, photoresponse of the laser-patterned device at sub-band photon energy (at 1250 nm) can be characterized by a relatively sharp, yet clear sub-linear power dependence with exponential factor of 0.74, while the pristine Si p-n junction is completely insensitive to such illumination due to the absence of the defect states. %However, Si photo-detector with spiky morphology still possesses relatively sharp power dependent photoresponse curve despite defect states introduced. It is more likely to be that mentioned above periods matching is responsible for a certain gain of the resulted photoresponse under sub-band gap illumination.

\section{Conclusions}
Here, we demonstrated how simultaneous action of electromagnetic and hydrodynamic processes driven by ultrashort multi-pulse laser exposure at the monocrystalline Si interface can be used to control the resulting micro- and nanoscale surface nanomorphology spanning from the regular LIPSSs to quasi-ordered (or randomly arranged) spikes. Our studies clearly showed that the electromagnetic processes (such as polarization-dependent excitation and interference of the surface waves) dominating at initial steps of the morphology evolution define the onset of the hydrodynamic processes and resulting morphology upon promoted multi-pulse exposure. This results in appearance of the quasi-regular morphologies having the surface features defined by the hydrodynamic instability and oriented parallel to the polarization vector of the incident radiation. Our studies add an important brick into fundamental understanding of the interaction of ultrashort laser pulses with the silicon that has been used for a long time as a key benchmark material for such studies. In particular, our results clarify the origin of formation of widely reported supra-wavelength surface features following polarization direction of the incident radiation.

From application point of view, our work highlights practical benefits of liquid-assisted laser processing of Si. In a sharp contrast to the well studied processing of Si under ambient environment, methanol not only reduces the efficiency of Si oxidation and removes debris from the nanotextured surface without any post-processing steps but also facilitates formation of the regular plasmon-mediated subwavelength LIPSSs with the periodicity $\approx\lambda/2$ as well as denser quasi-ordered spiky features. Considering wide applicability of silicon in optoelectronics, all-dielectric photonics and optical sensors, such advanced maskless and straightforward nanotexturing procedure is expected to be of high demand for future applications. As a good demonstration of the practical applicability of the developed approach, using direct laser patterning without any post-annealing and cleaning steps we realized the self-powered vertical p-n junction Si photodetector with twice larger photoresponse at sub-bandgap photon energies and expanded NIR operation range.

\section{Methods}

\subsection{Laser patterning}
Monocrystalline Si wafers were placed into a quartz cuvette filled with a methanol and directly processed with a second-harmonic (515 nm) fs-laser (200 fs) pulse generated at 1 kHz pulse repetition rate by a regeneratively amplified laser system (Pharos, Light Conversion). For certain experiments, as-generated linearly polarized Gaussian-shape laser beam was shaped into a CV beam by passing through a radial polarization converter ($S$-waveplate; \cite{beresna2011radially}). The laser texturing was performed through the static 4-mm thick layer of isopropanol with a Gaussian/CV beam focused onto the sample surface with a dry microscope objective having numerical aperture (NA) of 0.13. The size of the focal spot $D_{th}$ in the case of Gaussian-shape beam was evaluated using Liu method \cite{liu1982simple}. More specifically, we plotted the squared diameter of the single-pulse Si surface modification as a function of the incident pulse energy in logarithmic scale $D^2$-ln[$E$]. The beam size $D_{th}\approx$5 $\mu$m was obtained from the slope of the linear fit of the obtained dependence. In the case of CV beam generated by the $S$-waveplate, we used the Lambert function $W$ to calculate the inner and outer radii of the focal-plane annular distribution $\cite{zhang2017dimensional}$:

\begin{equation}\label{Eq:areas}
 r_{inner} = \sqrt{\frac{m}{2}} \sqrt{-W \left( 0, -\frac{\left( e^{-m-2} m^{m}\right)^ {1/m}}{m} \right) } \omega,
\end{equation}

and

\begin{equation}\label{Eq:areas}
r_{outer} = \sqrt{\frac{m}{2}} \sqrt{-W \left( -1, -\frac{\left( e^{-m-2} m^{m}\right)^ {1/m}}{m} \right) } \omega
\end{equation}

Then, the ratio of the areas of the Gaussian beam $S_{gauss}$=$\pi D^2_{th}/4$ and the CV beam $S_{CV} = \pi\left(r_{outer}^{2} - r_{inner}^{2} \right)$ for the focusing conditions ($NA$=0.13) can be assessed to be $S_{CV}/S_{gauss} = $2.2.

\subsection{Characterization}
Structural and morphological features of the laser-patterned surfaces were characterized using SEM device (Ultra 55+, Carl Zeiss) equipped with an EDX modality for chemical analysis. TEM studies (Titan 60-300, Thermo Fisher) were carried out with a thin lamella prepared using using dual-beam (electron/ion) device (Helios 450, Thermo Fisher). Preparation steps include (i) coverage of the sight of interest with a protective Pt layer, (ii) FIB cutting and thinning down to electron transparency and (iii) lamella extraction for TEM imaging.

Micro-Raman spectroscopy (Integra Spectra II, NT-MDT) was carried out with a CW laser pump centered at 473 nm. Linearly polarized laser radiation was focused onto the sample surface with a dry microscope objective having NA=0.28. Raman spectra were averaged over at least 20 surface sites for each laser-patterned area.

Fast Fourier Transform (FFT) analysis of the SEM images of the LIPSSs and spiky Si produced at variable irradiation conditions was performed with an ImageJ package. The same software was applied to evaluate average density of the self-organized structures using a find maxima tool.

Finite-difference time domain simulations were carried out to calculate the absorption (E$^2$Im[$\epsilon$], where E is an EM field amplitude and Im[$\epsilon$] is the dielectric permittivity of Si) of the incident laser pulses by the LIPSSs with variable morphologies. In particular, we considered the smooth Si surface having several pitches with a semi-elliptical profile (depth of 200 nm and width of 150 nm) arranged at the period of 260 nm that is fully consistent with the experimental data. Destabilizations of the grating morphology were modeled either as a local elliptical-shape protrusions decorating the ridges or larger protrusions filling the space between neighboring ridges (Fig. 1e). Si grating was placed in the background medium with a refractive index of 1.33 corresponding to those for methanol \cite{moutzouris2014refractive} and exposed from the top by a Gaussian-shape (1/e-diameter of 3 $\mu$m) pulse having the polarization perpendicular with respect to the ridge orientation. Computational volume was limited by perfectly matched layers.

Reflectivity of the laser-patterned surface in the visible (0.5 - 0.7 $\mu$m) and SWIR (1 - 3 $\mu$m) spectral ranges was assessed using an integrated-sphere spectrometer (Cary 5000, Varian) and Fourier-Transform IR spectrometer (Vertex 80, Bruker) coupled to the IR microscope (Hyperion 2000, Bruker), respectively.

\subsection{Fabrication and characterization of a Si photo-detector with a vertical p-n junction}
Monocrystalline n-type (resistivity of 0.1 Ohm$\cdot$cm) Si wafer with a lateral size of 6$\times$6 mm$^2$ with an epitaxially grown 2 $\mu$m thick p-type Si top layer (resistivity of 10 Ohm$\cdot$cm) was laser-textured following the previously described procedure to create a destabilized LIPSS morphology within certain surface areas (Fig. 6a,b). Then, 200 nm thick Al film was magnetron sputtered through the shadow mask to create a patched square-loop top electrode. Similarly, 2 nm of Sb film covered with 100 nm thick Au layer was deposited to form a filled square bottom electrode. Resulting device was annealed in vacuum (base pressure of 5$\times$10$^{-6}$ Torr) for 30 min at 400$^\circ$C to improve surface adhesion and reduce contact resistance. Same procedure was used to produce non-patterned (pristine) reference photodetector. Pristine and patterned Si photo-detectors were chip-packaged using ultrasonic welding with Al microwires.

The photo-responsivity of the Si p-n junction devices was probed under room tempereture and zero bias voltage using spectrally tunable laser radiation generated at 80 MHz pulse repetition rate by a femtosecond optical parametric oscillator (TOPOL, Avesta Project LTD.). The output laser beam was shaped by a lens to yield a pump spot diameter $\approx$ 60 $\mu$m.  Different sites of the laser-patterned and pristine devices were probed for statistical averaging of the results. The incident power at the selected wavelength within the spectral range from 700 to 1300 nm was measured by a Ge optical power meter (Newport). The photocurrent was directly captured by a Keithley sourcemeter from the illuminated devices. Photoresponse was calculated as the generated photocurrent divided by the power of incident light. Attenuated laser lines from the TOPOL setup were used to register the photoresponse $versus$ incident optical irradiation power for the patterned and pristine Si p-n junction photo-detectors.

%\section{Associated content}

%\section*{Author Contributions}

\section*{Conflict of interest
}
The authors declare no competing financial interests.

\section*{Acknowledgements}
This work was supported by the Russian Science Foundation (Grant no. 21-79-20075).


\begin{thebibliography}{75}%
\makeatletter
\providecommand \@ifxundefined [1]{%
 \@ifx{#1\undefined}
}%
\providecommand \@ifnum [1]{%
 \ifnum #1\expandafter \@firstoftwo
 \else \expandafter \@secondoftwo
 \fi
}%
\providecommand \@ifx [1]{%
 \ifx #1\expandafter \@firstoftwo
 \else \expandafter \@secondoftwo
 \fi
}%
\providecommand \natexlab [1]{#1}%
\providecommand \enquote  [1]{``#1''}%
\providecommand \bibnamefont  [1]{#1}%
\providecommand \bibfnamefont [1]{#1}%
\providecommand \citenamefont [1]{#1}%
\providecommand \href@noop [0]{\@secondoftwo}%
\providecommand \href [0]{\begingroup \@sanitize@url \@href}%
\providecommand \@href[1]{\@@startlink{#1}\@@href}%
\providecommand \@@href[1]{\endgroup#1\@@endlink}%
\providecommand \@sanitize@url [0]{\catcode `\\12\catcode `\$12\catcode
  `\&12\catcode `\#12\catcode `\^12\catcode `\_12\catcode `\%12\relax}%
\providecommand \@@startlink[1]{}%
\providecommand \@@endlink[0]{}%
\providecommand \url  [0]{\begingroup\@sanitize@url \@url }%
\providecommand \@url [1]{\endgroup\@href {#1}{\urlprefix }}%
\providecommand \urlprefix  [0]{URL }%
\providecommand \Eprint [0]{\href }%
\providecommand \doibase [0]{http://dx.doi.org/}%
\providecommand \selectlanguage [0]{\@gobble}%
\providecommand \bibinfo  [0]{\@secondoftwo}%
\providecommand \bibfield  [0]{\@secondoftwo}%
\providecommand \translation [1]{[#1]}%
\providecommand \BibitemOpen [0]{}%
\providecommand \bibitemStop [0]{}%
\providecommand \bibitemNoStop [0]{.\EOS\space}%
\providecommand \EOS [0]{\spacefactor3000\relax}%
\providecommand \BibitemShut  [1]{\csname bibitem#1\endcsname}%
\let\auto@bib@innerbib\@empty
%</preamble>
\bibitem [{\citenamefont {Sugioka}\ and\ \citenamefont
  {Cheng}(2014)}]{sugioka2014ultrafast}%
  \BibitemOpen
  \bibfield  {author} {\bibinfo {author} {\bibfnamefont {K.}~\bibnamefont
  {Sugioka}}\ and\ \bibinfo {author} {\bibfnamefont {Y.}~\bibnamefont
  {Cheng}},\ }\href@noop {} {\bibfield  {journal} {\bibinfo  {journal} {Light:
  Science \& Applications}\ }\textbf {\bibinfo {volume} {3}},\ \bibinfo {pages}
  {e149} (\bibinfo {year} {2014})}\BibitemShut {NoStop}%
\bibitem [{\citenamefont {Vorobyev}\ and\ \citenamefont
  {Guo}(2013)}]{vorobyev2013direct}%
  \BibitemOpen
  \bibfield  {author} {\bibinfo {author} {\bibfnamefont {A.~Y.}\ \bibnamefont
  {Vorobyev}}\ and\ \bibinfo {author} {\bibfnamefont {C.}~\bibnamefont {Guo}},\
  }\href@noop {} {\bibfield  {journal} {\bibinfo  {journal} {Laser \& Photonics
  Reviews}\ }\textbf {\bibinfo {volume} {7}},\ \bibinfo {pages} {385} (\bibinfo
  {year} {2013})}\BibitemShut {NoStop}%
\bibitem [{\citenamefont {Wei}\ \emph {et~al.}(2020)\citenamefont {Wei},
  \citenamefont {Zhang}, \citenamefont {Cheng}, \citenamefont {Sun},
  \citenamefont {Zhu},\ and\ \citenamefont {Li}}]{wei2020overview}%
  \BibitemOpen
  \bibfield  {author} {\bibinfo {author} {\bibfnamefont {C.}~\bibnamefont
  {Wei}}, \bibinfo {author} {\bibfnamefont {Z.}~\bibnamefont {Zhang}}, \bibinfo
  {author} {\bibfnamefont {D.}~\bibnamefont {Cheng}}, \bibinfo {author}
  {\bibfnamefont {Z.}~\bibnamefont {Sun}}, \bibinfo {author} {\bibfnamefont
  {M.}~\bibnamefont {Zhu}}, \ and\ \bibinfo {author} {\bibfnamefont
  {L.}~\bibnamefont {Li}},\ }\href@noop {} {\bibfield  {journal} {\bibinfo
  {journal} {International Journal of Extreme Manufacturing}\ }\textbf
  {\bibinfo {volume} {3}},\ \bibinfo {pages} {012003} (\bibinfo {year}
  {2020})}\BibitemShut {NoStop}%
\bibitem [{\citenamefont {Stratakis}\ \emph {et~al.}(2020)\citenamefont
  {Stratakis}, \citenamefont {Bonse}, \citenamefont {Heitz}, \citenamefont
  {Siegel}, \citenamefont {Tsibidis}, \citenamefont {Skoulas}, \citenamefont
  {Papadopoulos}, \citenamefont {Mimidis}, \citenamefont {Joel}, \citenamefont
  {Comanns}, \citenamefont {Kruger}, \citenamefont {Florian}, \citenamefont
  {Fuentes-Edfuf}, \citenamefont {Solis},\ and\ \citenamefont
  {Baumgartnerh}}]{stratakis2020laser}%
  \BibitemOpen
  \bibfield  {author} {\bibinfo {author} {\bibfnamefont {E.}~\bibnamefont
  {Stratakis}}, \bibinfo {author} {\bibfnamefont {J.}~\bibnamefont {Bonse}},
  \bibinfo {author} {\bibfnamefont {J.}~\bibnamefont {Heitz}}, \bibinfo
  {author} {\bibfnamefont {J.}~\bibnamefont {Siegel}}, \bibinfo {author}
  {\bibfnamefont {G.}~\bibnamefont {Tsibidis}}, \bibinfo {author}
  {\bibfnamefont {E.}~\bibnamefont {Skoulas}}, \bibinfo {author} {\bibfnamefont
  {A.}~\bibnamefont {Papadopoulos}}, \bibinfo {author} {\bibfnamefont
  {A.}~\bibnamefont {Mimidis}}, \bibinfo {author} {\bibfnamefont {A.-C.}\
  \bibnamefont {Joel}}, \bibinfo {author} {\bibfnamefont {P.}~\bibnamefont
  {Comanns}}, \bibinfo {author} {\bibfnamefont {J.}~\bibnamefont {Kruger}},
  \bibinfo {author} {\bibfnamefont {C.}~\bibnamefont {Florian}}, \bibinfo
  {author} {\bibfnamefont {Y.}~\bibnamefont {Fuentes-Edfuf}}, \bibinfo {author}
  {\bibfnamefont {J.}~\bibnamefont {Solis}}, \ and\ \bibinfo {author}
  {\bibfnamefont {W.}~\bibnamefont {Baumgartnerh}},\ }\href@noop {} {\bibfield
  {journal} {\bibinfo  {journal} {Materials Science and Engineering: R:
  Reports}\ }\textbf {\bibinfo {volume} {141}},\ \bibinfo {pages} {100562}
  (\bibinfo {year} {2020})}\BibitemShut {NoStop}%
\bibitem [{\citenamefont {Dalloz}\ \emph {et~al.}(2022)\citenamefont {Dalloz},
  \citenamefont {Le}, \citenamefont {Hebert}, \citenamefont {Eles},
  \citenamefont {Flores~Figueroa}, \citenamefont {Hubert}, \citenamefont {Ma},
  \citenamefont {Sharma}, \citenamefont {Vocanson}, \citenamefont {Ayala} \emph
  {et~al.}}]{dalloz2022anti}%
  \BibitemOpen
  \bibfield  {author} {\bibinfo {author} {\bibfnamefont {N.}~\bibnamefont
  {Dalloz}}, \bibinfo {author} {\bibfnamefont {V.~D.}\ \bibnamefont {Le}},
  \bibinfo {author} {\bibfnamefont {M.}~\bibnamefont {Hebert}}, \bibinfo
  {author} {\bibfnamefont {B.}~\bibnamefont {Eles}}, \bibinfo {author}
  {\bibfnamefont {M.~A.}\ \bibnamefont {Flores~Figueroa}}, \bibinfo {author}
  {\bibfnamefont {C.}~\bibnamefont {Hubert}}, \bibinfo {author} {\bibfnamefont
  {H.}~\bibnamefont {Ma}}, \bibinfo {author} {\bibfnamefont {N.}~\bibnamefont
  {Sharma}}, \bibinfo {author} {\bibfnamefont {F.}~\bibnamefont {Vocanson}},
  \bibinfo {author} {\bibfnamefont {S.}~\bibnamefont {Ayala}},  \emph
  {et~al.},\ }\href@noop {} {\bibfield  {journal} {\bibinfo  {journal}
  {Advanced Materials}\ }\textbf {\bibinfo {volume} {34}},\ \bibinfo {pages}
  {2104054} (\bibinfo {year} {2022})}\BibitemShut {NoStop}%
\bibitem [{\citenamefont {Porfirev}\ \emph {et~al.}(2023)\citenamefont
  {Porfirev}, \citenamefont {Khonina},\ and\ \citenamefont
  {Kuchmizhak}}]{porfirev2023light}%
  \BibitemOpen
  \bibfield  {author} {\bibinfo {author} {\bibfnamefont {A.}~\bibnamefont
  {Porfirev}}, \bibinfo {author} {\bibfnamefont {S.}~\bibnamefont {Khonina}}, \
  and\ \bibinfo {author} {\bibfnamefont {A.}~\bibnamefont {Kuchmizhak}},\
  }\href@noop {} {\bibfield  {journal} {\bibinfo  {journal} {Progress in
  Quantum Electronics}\ ,\ \bibinfo {pages} {100459}} (\bibinfo {year}
  {2023})}\BibitemShut {NoStop}%
\bibitem [{\citenamefont {Brandao}\ \emph {et~al.}(2023)\citenamefont
  {Brandao}, \citenamefont {Nakhoul}, \citenamefont {Duffner}, \citenamefont
  {Emonet}, \citenamefont {Garrelie}, \citenamefont {Habrard}, \citenamefont
  {Jacquenet}, \citenamefont {Pigeon}, \citenamefont {Sebban},\ and\
  \citenamefont {Colombier}}]{brandao2023learning}%
  \BibitemOpen
  \bibfield  {author} {\bibinfo {author} {\bibfnamefont {E.}~\bibnamefont
  {Brandao}}, \bibinfo {author} {\bibfnamefont {A.}~\bibnamefont {Nakhoul}},
  \bibinfo {author} {\bibfnamefont {S.}~\bibnamefont {Duffner}}, \bibinfo
  {author} {\bibfnamefont {R.}~\bibnamefont {Emonet}}, \bibinfo {author}
  {\bibfnamefont {F.}~\bibnamefont {Garrelie}}, \bibinfo {author}
  {\bibfnamefont {A.}~\bibnamefont {Habrard}}, \bibinfo {author} {\bibfnamefont
  {F.}~\bibnamefont {Jacquenet}}, \bibinfo {author} {\bibfnamefont
  {F.}~\bibnamefont {Pigeon}}, \bibinfo {author} {\bibfnamefont
  {M.}~\bibnamefont {Sebban}}, \ and\ \bibinfo {author} {\bibfnamefont {J.-P.}\
  \bibnamefont {Colombier}},\ }\href@noop {} {\bibfield  {journal} {\bibinfo
  {journal} {Physical Review Letters}\ }\textbf {\bibinfo {volume} {130}},\
  \bibinfo {pages} {226201} (\bibinfo {year} {2023})}\BibitemShut {NoStop}%
\bibitem [{\citenamefont {Stoian}\ and\ \citenamefont
  {Colombier}(2020)}]{stoian2020advances}%
  \BibitemOpen
  \bibfield  {author} {\bibinfo {author} {\bibfnamefont {R.}~\bibnamefont
  {Stoian}}\ and\ \bibinfo {author} {\bibfnamefont {J.-P.}\ \bibnamefont
  {Colombier}},\ }\href@noop {} {\bibfield  {journal} {\bibinfo  {journal}
  {Nanophotonics}\ }\textbf {\bibinfo {volume} {9}},\ \bibinfo {pages} {4665}
  (\bibinfo {year} {2020})}\BibitemShut {NoStop}%
\bibitem [{\citenamefont {Stoian}\ and\ \citenamefont
  {Bonse}(2023)}]{stoian2023ultrafast}%
  \BibitemOpen
  \bibfield  {author} {\bibinfo {author} {\bibfnamefont {R.}~\bibnamefont
  {Stoian}}\ and\ \bibinfo {author} {\bibfnamefont {J.}~\bibnamefont {Bonse}},\
  }\href@noop {} {\emph {\bibinfo {title} {Ultrafast laser nanostructuring: the
  pursuit of extreme scales}}},\ Vol.\ \bibinfo {volume} {239}\ (\bibinfo
  {publisher} {Springer Nature},\ \bibinfo {year} {2023})\BibitemShut {NoStop}%
\bibitem [{\citenamefont {Birnbaum}(1965)}]{birnbaum1965semiconductor}%
  \BibitemOpen
  \bibfield  {author} {\bibinfo {author} {\bibfnamefont {M.}~\bibnamefont
  {Birnbaum}},\ }\href@noop {} {\bibfield  {journal} {\bibinfo  {journal}
  {Journal of Applied Physics}\ }\textbf {\bibinfo {volume} {36}},\ \bibinfo
  {pages} {3688} (\bibinfo {year} {1965})}\BibitemShut {NoStop}%
\bibitem [{\citenamefont {Nemanich}\ \emph {et~al.}(1983)\citenamefont
  {Nemanich}, \citenamefont {Biegelsen},\ and\ \citenamefont
  {Hawkins}}]{nemanich1983aligned}%
  \BibitemOpen
  \bibfield  {author} {\bibinfo {author} {\bibfnamefont {R.}~\bibnamefont
  {Nemanich}}, \bibinfo {author} {\bibfnamefont {D.}~\bibnamefont {Biegelsen}},
  \ and\ \bibinfo {author} {\bibfnamefont {W.}~\bibnamefont {Hawkins}},\
  }\href@noop {} {\bibfield  {journal} {\bibinfo  {journal} {Physical Review
  B}\ }\textbf {\bibinfo {volume} {27}},\ \bibinfo {pages} {7817} (\bibinfo
  {year} {1983})}\BibitemShut {NoStop}%
\bibitem [{\citenamefont {Buividas}\ \emph {et~al.}(2014)\citenamefont
  {Buividas}, \citenamefont {Mikutis},\ and\ \citenamefont
  {Juodkazis}}]{buividas2014surface}%
  \BibitemOpen
  \bibfield  {author} {\bibinfo {author} {\bibfnamefont {R.}~\bibnamefont
  {Buividas}}, \bibinfo {author} {\bibfnamefont {M.}~\bibnamefont {Mikutis}}, \
  and\ \bibinfo {author} {\bibfnamefont {S.}~\bibnamefont {Juodkazis}},\
  }\href@noop {} {\bibfield  {journal} {\bibinfo  {journal} {Progress in
  Quantum Electronics}\ }\textbf {\bibinfo {volume} {38}},\ \bibinfo {pages}
  {119} (\bibinfo {year} {2014})}\BibitemShut {NoStop}%
\bibitem [{\citenamefont {Bonse}\ \emph {et~al.}(2016)\citenamefont {Bonse},
  \citenamefont {H{\"o}hm}, \citenamefont {Kirner}, \citenamefont {Rosenfeld},\
  and\ \citenamefont {Kr{\"u}ger}}]{bonse2016laser}%
  \BibitemOpen
  \bibfield  {author} {\bibinfo {author} {\bibfnamefont {J.}~\bibnamefont
  {Bonse}}, \bibinfo {author} {\bibfnamefont {S.}~\bibnamefont {H{\"o}hm}},
  \bibinfo {author} {\bibfnamefont {S.~V.}\ \bibnamefont {Kirner}}, \bibinfo
  {author} {\bibfnamefont {A.}~\bibnamefont {Rosenfeld}}, \ and\ \bibinfo
  {author} {\bibfnamefont {J.}~\bibnamefont {Kr{\"u}ger}},\ }\href@noop {}
  {\bibfield  {journal} {\bibinfo  {journal} {IEEE Journal of selected topics
  in quantum electronics}\ }\textbf {\bibinfo {volume} {23}} (\bibinfo {year}
  {2016})}\BibitemShut {NoStop}%
\bibitem [{\citenamefont {Bonse}\ and\ \citenamefont
  {Gr{\"a}f}(2020)}]{bonse2020maxwell}%
  \BibitemOpen
  \bibfield  {author} {\bibinfo {author} {\bibfnamefont {J.}~\bibnamefont
  {Bonse}}\ and\ \bibinfo {author} {\bibfnamefont {S.}~\bibnamefont
  {Gr{\"a}f}},\ }\href@noop {} {\bibfield  {journal} {\bibinfo  {journal}
  {Laser \& Photonics Reviews}\ }\textbf {\bibinfo {volume} {14}},\ \bibinfo
  {pages} {2000215} (\bibinfo {year} {2020})}\BibitemShut {NoStop}%
\bibitem [{\citenamefont {Tsibidis}\ \emph {et~al.}(2016)\citenamefont
  {Tsibidis}, \citenamefont {Skoulas}, \citenamefont {Papadopoulos},\ and\
  \citenamefont {Stratakis}}]{tsibidis2016convection}%
  \BibitemOpen
  \bibfield  {author} {\bibinfo {author} {\bibfnamefont {G.~D.}\ \bibnamefont
  {Tsibidis}}, \bibinfo {author} {\bibfnamefont {E.}~\bibnamefont {Skoulas}},
  \bibinfo {author} {\bibfnamefont {A.}~\bibnamefont {Papadopoulos}}, \ and\
  \bibinfo {author} {\bibfnamefont {E.}~\bibnamefont {Stratakis}},\ }\href@noop
  {} {\bibfield  {journal} {\bibinfo  {journal} {Physical Review B}\ }\textbf
  {\bibinfo {volume} {94}},\ \bibinfo {pages} {081305} (\bibinfo {year}
  {2016})}\BibitemShut {NoStop}%
\bibitem [{\citenamefont {Xu}\ \emph {et~al.}(2019)\citenamefont {Xu},
  \citenamefont {Sun}, \citenamefont {Yao}, \citenamefont {Liu}, \citenamefont
  {Miao}, \citenamefont {Jiang}, \citenamefont {Wang}, \citenamefont {Jiang},
  \citenamefont {Yuan},\ and\ \citenamefont {Zu}}]{xu2019periodic}%
  \BibitemOpen
  \bibfield  {author} {\bibinfo {author} {\bibfnamefont {S.-Z.}\ \bibnamefont
  {Xu}}, \bibinfo {author} {\bibfnamefont {K.}~\bibnamefont {Sun}}, \bibinfo
  {author} {\bibfnamefont {C.-Z.}\ \bibnamefont {Yao}}, \bibinfo {author}
  {\bibfnamefont {H.}~\bibnamefont {Liu}}, \bibinfo {author} {\bibfnamefont
  {X.-X.}\ \bibnamefont {Miao}}, \bibinfo {author} {\bibfnamefont {Y.-L.}\
  \bibnamefont {Jiang}}, \bibinfo {author} {\bibfnamefont {H.-J.}\ \bibnamefont
  {Wang}}, \bibinfo {author} {\bibfnamefont {X.-D.}\ \bibnamefont {Jiang}},
  \bibinfo {author} {\bibfnamefont {X.-D.}\ \bibnamefont {Yuan}}, \ and\
  \bibinfo {author} {\bibfnamefont {X.-T.}\ \bibnamefont {Zu}},\ }\href@noop {}
  {\bibfield  {journal} {\bibinfo  {journal} {Optics Express}\ }\textbf
  {\bibinfo {volume} {27}},\ \bibinfo {pages} {8983} (\bibinfo {year}
  {2019})}\BibitemShut {NoStop}%
\bibitem [{\citenamefont {Zhang}\ \emph {et~al.}(2021)\citenamefont {Zhang},
  \citenamefont {Liu},\ and\ \citenamefont {Li}}]{zhang2021irregular}%
  \BibitemOpen
  \bibfield  {author} {\bibinfo {author} {\bibfnamefont {D.}~\bibnamefont
  {Zhang}}, \bibinfo {author} {\bibfnamefont {R.}~\bibnamefont {Liu}}, \ and\
  \bibinfo {author} {\bibfnamefont {Z.}~\bibnamefont {Li}},\ }\href@noop {}
  {\bibfield  {journal} {\bibinfo  {journal} {International Journal of Extreme
  Manufacturing}\ }\textbf {\bibinfo {volume} {4}},\ \bibinfo {pages} {015102}
  (\bibinfo {year} {2021})}\BibitemShut {NoStop}%
\bibitem [{\citenamefont {Green}\ \emph {et~al.}(2021)\citenamefont {Green},
  \citenamefont {Dunlop}, \citenamefont {Hohl-Ebinger}, \citenamefont
  {Yoshita}, \citenamefont {Kopidakis},\ and\ \citenamefont
  {Hao}}]{green2021solar}%
  \BibitemOpen
  \bibfield  {author} {\bibinfo {author} {\bibfnamefont {M.}~\bibnamefont
  {Green}}, \bibinfo {author} {\bibfnamefont {E.}~\bibnamefont {Dunlop}},
  \bibinfo {author} {\bibfnamefont {J.}~\bibnamefont {Hohl-Ebinger}}, \bibinfo
  {author} {\bibfnamefont {M.}~\bibnamefont {Yoshita}}, \bibinfo {author}
  {\bibfnamefont {N.}~\bibnamefont {Kopidakis}}, \ and\ \bibinfo {author}
  {\bibfnamefont {X.}~\bibnamefont {Hao}},\ }\href@noop {} {\bibfield
  {journal} {\bibinfo  {journal} {Progress in photovoltaics: research and
  applications}\ }\textbf {\bibinfo {volume} {29}},\ \bibinfo {pages} {3}
  (\bibinfo {year} {2021})}\BibitemShut {NoStop}%
\bibitem [{\citenamefont {Baranov}\ \emph {et~al.}(2017)\citenamefont
  {Baranov}, \citenamefont {Zuev}, \citenamefont {Lepeshov}, \citenamefont
  {Kotov}, \citenamefont {Krasnok}, \citenamefont {Evlyukhin},\ and\
  \citenamefont {Chichkov}}]{baranov2017all}%
  \BibitemOpen
  \bibfield  {author} {\bibinfo {author} {\bibfnamefont {D.~G.}\ \bibnamefont
  {Baranov}}, \bibinfo {author} {\bibfnamefont {D.~A.}\ \bibnamefont {Zuev}},
  \bibinfo {author} {\bibfnamefont {S.~I.}\ \bibnamefont {Lepeshov}}, \bibinfo
  {author} {\bibfnamefont {O.~V.}\ \bibnamefont {Kotov}}, \bibinfo {author}
  {\bibfnamefont {A.~E.}\ \bibnamefont {Krasnok}}, \bibinfo {author}
  {\bibfnamefont {A.~B.}\ \bibnamefont {Evlyukhin}}, \ and\ \bibinfo {author}
  {\bibfnamefont {B.~N.}\ \bibnamefont {Chichkov}},\ }\href@noop {} {\bibfield
  {journal} {\bibinfo  {journal} {Optica}\ }\textbf {\bibinfo {volume} {4}},\
  \bibinfo {pages} {814} (\bibinfo {year} {2017})}\BibitemShut {NoStop}%
\bibitem [{\citenamefont {Le~Harzic}\ \emph {et~al.}(2005)\citenamefont
  {Le~Harzic}, \citenamefont {Schuck}, \citenamefont {Sauer}, \citenamefont
  {Anhut}, \citenamefont {Riemann},\ and\ \citenamefont
  {K{\"o}nig}}]{le2005sub}%
  \BibitemOpen
  \bibfield  {author} {\bibinfo {author} {\bibfnamefont {R.}~\bibnamefont
  {Le~Harzic}}, \bibinfo {author} {\bibfnamefont {H.}~\bibnamefont {Schuck}},
  \bibinfo {author} {\bibfnamefont {D.}~\bibnamefont {Sauer}}, \bibinfo
  {author} {\bibfnamefont {T.}~\bibnamefont {Anhut}}, \bibinfo {author}
  {\bibfnamefont {I.}~\bibnamefont {Riemann}}, \ and\ \bibinfo {author}
  {\bibfnamefont {K.}~\bibnamefont {K{\"o}nig}},\ }\href@noop {} {\bibfield
  {journal} {\bibinfo  {journal} {Optics Express}\ }\textbf {\bibinfo {volume}
  {13}},\ \bibinfo {pages} {6651} (\bibinfo {year} {2005})}\BibitemShut
  {NoStop}%
\bibitem [{\citenamefont {Shen}\ \emph {et~al.}(2008)\citenamefont {Shen},
  \citenamefont {Carey}, \citenamefont {Crouch}, \citenamefont {Kandyla},
  \citenamefont {Stone},\ and\ \citenamefont {Mazur}}]{shen2008high}%
  \BibitemOpen
  \bibfield  {author} {\bibinfo {author} {\bibfnamefont {M.}~\bibnamefont
  {Shen}}, \bibinfo {author} {\bibfnamefont {J.~E.}\ \bibnamefont {Carey}},
  \bibinfo {author} {\bibfnamefont {C.~H.}\ \bibnamefont {Crouch}}, \bibinfo
  {author} {\bibfnamefont {M.}~\bibnamefont {Kandyla}}, \bibinfo {author}
  {\bibfnamefont {H.~A.}\ \bibnamefont {Stone}}, \ and\ \bibinfo {author}
  {\bibfnamefont {E.}~\bibnamefont {Mazur}},\ }\href@noop {} {\bibfield
  {journal} {\bibinfo  {journal} {Nano letters}\ }\textbf {\bibinfo {volume}
  {8}},\ \bibinfo {pages} {2087} (\bibinfo {year} {2008})}\BibitemShut
  {NoStop}%
\bibitem [{\citenamefont {Le~Harzic}\ \emph {et~al.}(2011)\citenamefont
  {Le~Harzic}, \citenamefont {D{\"o}rr}, \citenamefont {Sauer}, \citenamefont
  {Stracke},\ and\ \citenamefont {Zimmermann}}]{le2011generation}%
  \BibitemOpen
  \bibfield  {author} {\bibinfo {author} {\bibfnamefont {R.}~\bibnamefont
  {Le~Harzic}}, \bibinfo {author} {\bibfnamefont {D.}~\bibnamefont {D{\"o}rr}},
  \bibinfo {author} {\bibfnamefont {D.}~\bibnamefont {Sauer}}, \bibinfo
  {author} {\bibfnamefont {F.}~\bibnamefont {Stracke}}, \ and\ \bibinfo
  {author} {\bibfnamefont {H.}~\bibnamefont {Zimmermann}},\ }\href@noop {}
  {\bibfield  {journal} {\bibinfo  {journal} {Applied Physics Letters}\
  }\textbf {\bibinfo {volume} {98}},\ \bibinfo {pages} {211905} (\bibinfo
  {year} {2011})}\BibitemShut {NoStop}%
\bibitem [{\citenamefont {Straub}\ \emph {et~al.}(2012)\citenamefont {Straub},
  \citenamefont {Afshar}, \citenamefont {Feili}, \citenamefont {Seidel},\ and\
  \citenamefont {K{\"o}nig}}]{straub2012periodic}%
  \BibitemOpen
  \bibfield  {author} {\bibinfo {author} {\bibfnamefont {M.}~\bibnamefont
  {Straub}}, \bibinfo {author} {\bibfnamefont {M.}~\bibnamefont {Afshar}},
  \bibinfo {author} {\bibfnamefont {D.}~\bibnamefont {Feili}}, \bibinfo
  {author} {\bibfnamefont {H.}~\bibnamefont {Seidel}}, \ and\ \bibinfo {author}
  {\bibfnamefont {K.}~\bibnamefont {K{\"o}nig}},\ }\href@noop {} {\bibfield
  {journal} {\bibinfo  {journal} {Optics letters}\ }\textbf {\bibinfo {volume}
  {37}},\ \bibinfo {pages} {190} (\bibinfo {year} {2012})}\BibitemShut
  {NoStop}%
\bibitem [{\citenamefont {Hamad}\ \emph {et~al.}(2014)\citenamefont {Hamad},
  \citenamefont {Podagatlapalli}, \citenamefont {Vendamani}, \citenamefont
  {Nageswara~Rao}, \citenamefont {Pathak}, \citenamefont {Tewari},\ and\
  \citenamefont {Venugopal~Rao}}]{hamad2014femtosecond}%
  \BibitemOpen
  \bibfield  {author} {\bibinfo {author} {\bibfnamefont {S.}~\bibnamefont
  {Hamad}}, \bibinfo {author} {\bibfnamefont {G.~K.}\ \bibnamefont
  {Podagatlapalli}}, \bibinfo {author} {\bibfnamefont {V.}~\bibnamefont
  {Vendamani}}, \bibinfo {author} {\bibfnamefont {S.}~\bibnamefont
  {Nageswara~Rao}}, \bibinfo {author} {\bibfnamefont {A.}~\bibnamefont
  {Pathak}}, \bibinfo {author} {\bibfnamefont {S.~P.}\ \bibnamefont {Tewari}},
  \ and\ \bibinfo {author} {\bibfnamefont {S.}~\bibnamefont {Venugopal~Rao}},\
  }\href@noop {} {\bibfield  {journal} {\bibinfo  {journal} {The Journal of
  Physical Chemistry C}\ }\textbf {\bibinfo {volume} {118}},\ \bibinfo {pages}
  {7139} (\bibinfo {year} {2014})}\BibitemShut {NoStop}%
\bibitem [{\citenamefont {Meng}\ \emph {et~al.}(2017)\citenamefont {Meng},
  \citenamefont {Jiang}, \citenamefont {Li}, \citenamefont {Xu}, \citenamefont
  {Shi}, \citenamefont {Yan},\ and\ \citenamefont {Lu}}]{meng2017dual}%
  \BibitemOpen
  \bibfield  {author} {\bibinfo {author} {\bibfnamefont {G.}~\bibnamefont
  {Meng}}, \bibinfo {author} {\bibfnamefont {L.}~\bibnamefont {Jiang}},
  \bibinfo {author} {\bibfnamefont {X.}~\bibnamefont {Li}}, \bibinfo {author}
  {\bibfnamefont {Y.}~\bibnamefont {Xu}}, \bibinfo {author} {\bibfnamefont
  {X.}~\bibnamefont {Shi}}, \bibinfo {author} {\bibfnamefont {R.}~\bibnamefont
  {Yan}}, \ and\ \bibinfo {author} {\bibfnamefont {Y.}~\bibnamefont {Lu}},\
  }\href@noop {} {\bibfield  {journal} {\bibinfo  {journal} {Applied Surface
  Science}\ }\textbf {\bibinfo {volume} {410}},\ \bibinfo {pages} {22}
  (\bibinfo {year} {2017})}\BibitemShut {NoStop}%
\bibitem [{\citenamefont {Yiannakou}\ \emph {et~al.}(2017)\citenamefont
  {Yiannakou}, \citenamefont {Simitzi}, \citenamefont {Manousaki},
  \citenamefont {Fotakis}, \citenamefont {Ranella},\ and\ \citenamefont
  {Stratakis}}]{yiannakou2017cell}%
  \BibitemOpen
  \bibfield  {author} {\bibinfo {author} {\bibfnamefont {C.}~\bibnamefont
  {Yiannakou}}, \bibinfo {author} {\bibfnamefont {C.}~\bibnamefont {Simitzi}},
  \bibinfo {author} {\bibfnamefont {A.}~\bibnamefont {Manousaki}}, \bibinfo
  {author} {\bibfnamefont {C.}~\bibnamefont {Fotakis}}, \bibinfo {author}
  {\bibfnamefont {A.}~\bibnamefont {Ranella}}, \ and\ \bibinfo {author}
  {\bibfnamefont {E.}~\bibnamefont {Stratakis}},\ }\href@noop {} {\bibfield
  {journal} {\bibinfo  {journal} {Biofabrication}\ }\textbf {\bibinfo {volume}
  {9}},\ \bibinfo {pages} {025024} (\bibinfo {year} {2017})}\BibitemShut
  {NoStop}%
\bibitem [{\citenamefont {Zhang}\ and\ \citenamefont
  {Sugioka}(2019)}]{zhang2019hierarchical}%
  \BibitemOpen
  \bibfield  {author} {\bibinfo {author} {\bibfnamefont {D.}~\bibnamefont
  {Zhang}}\ and\ \bibinfo {author} {\bibfnamefont {K.}~\bibnamefont
  {Sugioka}},\ }\href@noop {} {\bibfield  {journal} {\bibinfo  {journal}
  {Opto-Electronic Advances}\ }\textbf {\bibinfo {volume} {2}},\ \bibinfo
  {pages} {190002} (\bibinfo {year} {2019})}\BibitemShut {NoStop}%
\bibitem [{\citenamefont {Kesaev}\ \emph {et~al.}(2021)\citenamefont {Kesaev},
  \citenamefont {Nastulyavichus}, \citenamefont {Kudryashov}, \citenamefont
  {Kovalev}, \citenamefont {Stsepuro},\ and\ \citenamefont
  {Krasin}}]{kesaev2021nanopatterned}%
  \BibitemOpen
  \bibfield  {author} {\bibinfo {author} {\bibfnamefont {V.}~\bibnamefont
  {Kesaev}}, \bibinfo {author} {\bibfnamefont {A.}~\bibnamefont
  {Nastulyavichus}}, \bibinfo {author} {\bibfnamefont {S.}~\bibnamefont
  {Kudryashov}}, \bibinfo {author} {\bibfnamefont {M.}~\bibnamefont {Kovalev}},
  \bibinfo {author} {\bibfnamefont {N.}~\bibnamefont {Stsepuro}}, \ and\
  \bibinfo {author} {\bibfnamefont {G.}~\bibnamefont {Krasin}},\ }\href@noop {}
  {\bibfield  {journal} {\bibinfo  {journal} {Optical Materials Express}\
  }\textbf {\bibinfo {volume} {11}},\ \bibinfo {pages} {1971} (\bibinfo {year}
  {2021})}\BibitemShut {NoStop}%
\bibitem [{\citenamefont {Borodaenko}\ \emph {et~al.}(2021)\citenamefont
  {Borodaenko}, \citenamefont {Syubaev}, \citenamefont {Gurbatov},
  \citenamefont {Zhizhchenko}, \citenamefont {Porfirev}, \citenamefont
  {Khonina}, \citenamefont {Mitsai}, \citenamefont {Gerasimenko}, \citenamefont
  {Shevlyagin}, \citenamefont {Modin} \emph {et~al.}}]{borodaenko2021deep}%
  \BibitemOpen
  \bibfield  {author} {\bibinfo {author} {\bibfnamefont {Y.}~\bibnamefont
  {Borodaenko}}, \bibinfo {author} {\bibfnamefont {S.}~\bibnamefont {Syubaev}},
  \bibinfo {author} {\bibfnamefont {S.}~\bibnamefont {Gurbatov}}, \bibinfo
  {author} {\bibfnamefont {A.}~\bibnamefont {Zhizhchenko}}, \bibinfo {author}
  {\bibfnamefont {A.}~\bibnamefont {Porfirev}}, \bibinfo {author}
  {\bibfnamefont {S.}~\bibnamefont {Khonina}}, \bibinfo {author} {\bibfnamefont
  {E.}~\bibnamefont {Mitsai}}, \bibinfo {author} {\bibfnamefont {A.~V.}\
  \bibnamefont {Gerasimenko}}, \bibinfo {author} {\bibfnamefont
  {A.}~\bibnamefont {Shevlyagin}}, \bibinfo {author} {\bibfnamefont
  {E.}~\bibnamefont {Modin}},  \emph {et~al.},\ }\href@noop {} {\bibfield
  {journal} {\bibinfo  {journal} {ACS Applied Materials \& Interfaces}\
  }\textbf {\bibinfo {volume} {13}},\ \bibinfo {pages} {54551} (\bibinfo {year}
  {2021})}\BibitemShut {NoStop}%
\bibitem [{\citenamefont {Borodaenko}\ \emph {et~al.}(2022)\citenamefont
  {Borodaenko}, \citenamefont {Syubaev}, \citenamefont {Khairullina},
  \citenamefont {Tumkin}, \citenamefont {Gurbatov}, \citenamefont {Mironenko},
  \citenamefont {Mitsai}, \citenamefont {Zhizhchenko}, \citenamefont {Modin},
  \citenamefont {Gurevich} \emph {et~al.}}]{borodaenko2022demand}%
  \BibitemOpen
  \bibfield  {author} {\bibinfo {author} {\bibfnamefont {Y.}~\bibnamefont
  {Borodaenko}}, \bibinfo {author} {\bibfnamefont {S.}~\bibnamefont {Syubaev}},
  \bibinfo {author} {\bibfnamefont {E.}~\bibnamefont {Khairullina}}, \bibinfo
  {author} {\bibfnamefont {I.}~\bibnamefont {Tumkin}}, \bibinfo {author}
  {\bibfnamefont {S.}~\bibnamefont {Gurbatov}}, \bibinfo {author}
  {\bibfnamefont {A.}~\bibnamefont {Mironenko}}, \bibinfo {author}
  {\bibfnamefont {E.}~\bibnamefont {Mitsai}}, \bibinfo {author} {\bibfnamefont
  {A.}~\bibnamefont {Zhizhchenko}}, \bibinfo {author} {\bibfnamefont
  {E.}~\bibnamefont {Modin}}, \bibinfo {author} {\bibfnamefont {E.~L.}\
  \bibnamefont {Gurevich}},  \emph {et~al.},\ }\href@noop {} {\bibfield
  {journal} {\bibinfo  {journal} {Advanced Optical Materials}\ }\textbf
  {\bibinfo {volume} {10}},\ \bibinfo {pages} {2201094} (\bibinfo {year}
  {2022})}\BibitemShut {NoStop}%
\bibitem [{\citenamefont {Ma}\ \emph {et~al.}(2014)\citenamefont {Ma},
  \citenamefont {Si}, \citenamefont {Sun}, \citenamefont {Chen},\ and\
  \citenamefont {Hou}}]{ma2014progressive}%
  \BibitemOpen
  \bibfield  {author} {\bibinfo {author} {\bibfnamefont {Y.}~\bibnamefont
  {Ma}}, \bibinfo {author} {\bibfnamefont {J.}~\bibnamefont {Si}}, \bibinfo
  {author} {\bibfnamefont {X.}~\bibnamefont {Sun}}, \bibinfo {author}
  {\bibfnamefont {T.}~\bibnamefont {Chen}}, \ and\ \bibinfo {author}
  {\bibfnamefont {X.}~\bibnamefont {Hou}},\ }\href@noop {} {\bibfield
  {journal} {\bibinfo  {journal} {Applied surface science}\ }\textbf {\bibinfo
  {volume} {313}},\ \bibinfo {pages} {905} (\bibinfo {year}
  {2014})}\BibitemShut {NoStop}%
\bibitem [{\citenamefont {Zayarny}\ \emph {et~al.}(2016)\citenamefont
  {Zayarny}, \citenamefont {Ionin}, \citenamefont {Kudryashov}, \citenamefont
  {Makarov}, \citenamefont {Kuchmizhak}, \citenamefont {Vitrik},\ and\
  \citenamefont {Kulchin}}]{zayarny2016surface}%
  \BibitemOpen
  \bibfield  {author} {\bibinfo {author} {\bibfnamefont {D.}~\bibnamefont
  {Zayarny}}, \bibinfo {author} {\bibfnamefont {A.~A.}\ \bibnamefont {Ionin}},
  \bibinfo {author} {\bibfnamefont {S.~I.}\ \bibnamefont {Kudryashov}},
  \bibinfo {author} {\bibfnamefont {S.~V.}\ \bibnamefont {Makarov}}, \bibinfo
  {author} {\bibfnamefont {A.~A.}\ \bibnamefont {Kuchmizhak}}, \bibinfo
  {author} {\bibfnamefont {O.~B.}\ \bibnamefont {Vitrik}}, \ and\ \bibinfo
  {author} {\bibfnamefont {Y.~N.}\ \bibnamefont {Kulchin}},\ }\href@noop {}
  {\bibfield  {journal} {\bibinfo  {journal} {JETP Letters}\ }\textbf {\bibinfo
  {volume} {103}},\ \bibinfo {pages} {752} (\bibinfo {year}
  {2016})}\BibitemShut {NoStop}%
\bibitem [{\citenamefont {Nivas}\ \emph {et~al.}(2017)\citenamefont {Nivas},
  \citenamefont {He}, \citenamefont {Song}, \citenamefont {Rubano},
  \citenamefont {Vecchione}, \citenamefont {Paparo}, \citenamefont {Marrucci},
  \citenamefont {Bruzzese},\ and\ \citenamefont
  {Amoruso}}]{nivas2017femtosecond}%
  \BibitemOpen
  \bibfield  {author} {\bibinfo {author} {\bibfnamefont {J.~J.}\ \bibnamefont
  {Nivas}}, \bibinfo {author} {\bibfnamefont {S.}~\bibnamefont {He}}, \bibinfo
  {author} {\bibfnamefont {Z.}~\bibnamefont {Song}}, \bibinfo {author}
  {\bibfnamefont {A.}~\bibnamefont {Rubano}}, \bibinfo {author} {\bibfnamefont
  {A.}~\bibnamefont {Vecchione}}, \bibinfo {author} {\bibfnamefont
  {D.}~\bibnamefont {Paparo}}, \bibinfo {author} {\bibfnamefont
  {L.}~\bibnamefont {Marrucci}}, \bibinfo {author} {\bibfnamefont
  {R.}~\bibnamefont {Bruzzese}}, \ and\ \bibinfo {author} {\bibfnamefont
  {S.}~\bibnamefont {Amoruso}},\ }\href@noop {} {\bibfield  {journal} {\bibinfo
   {journal} {Applied Surface Science}\ }\textbf {\bibinfo {volume} {418}},\
  \bibinfo {pages} {565} (\bibinfo {year} {2017})}\BibitemShut {NoStop}%
\bibitem [{\citenamefont {Nivas}\ \emph {et~al.}(2018)\citenamefont {Nivas},
  \citenamefont {Anoop}, \citenamefont {Bruzzese}, \citenamefont {Philip},\
  and\ \citenamefont {Amoruso}}]{nivas2018direct}%
  \BibitemOpen
  \bibfield  {author} {\bibinfo {author} {\bibfnamefont {J.~J.}\ \bibnamefont
  {Nivas}}, \bibinfo {author} {\bibfnamefont {K.}~\bibnamefont {Anoop}},
  \bibinfo {author} {\bibfnamefont {R.}~\bibnamefont {Bruzzese}}, \bibinfo
  {author} {\bibfnamefont {R.}~\bibnamefont {Philip}}, \ and\ \bibinfo {author}
  {\bibfnamefont {S.}~\bibnamefont {Amoruso}},\ }\href@noop {} {\bibfield
  {journal} {\bibinfo  {journal} {Applied Physics Letters}\ }\textbf {\bibinfo
  {volume} {112}},\ \bibinfo {pages} {121601} (\bibinfo {year}
  {2018})}\BibitemShut {NoStop}%
\bibitem [{\citenamefont {Allahyari}\ \emph {et~al.}(2020)\citenamefont
  {Allahyari}, \citenamefont {Nivas}, \citenamefont {Skoulas}, \citenamefont
  {Bruzzese}, \citenamefont {Tsibidis}, \citenamefont {Stratakis},\ and\
  \citenamefont {Amoruso}}]{allahyari2020formation}%
  \BibitemOpen
  \bibfield  {author} {\bibinfo {author} {\bibfnamefont {E.}~\bibnamefont
  {Allahyari}}, \bibinfo {author} {\bibfnamefont {J.~J.}\ \bibnamefont
  {Nivas}}, \bibinfo {author} {\bibfnamefont {E.}~\bibnamefont {Skoulas}},
  \bibinfo {author} {\bibfnamefont {R.}~\bibnamefont {Bruzzese}}, \bibinfo
  {author} {\bibfnamefont {G.}~\bibnamefont {Tsibidis}}, \bibinfo {author}
  {\bibfnamefont {E.}~\bibnamefont {Stratakis}}, \ and\ \bibinfo {author}
  {\bibfnamefont {S.}~\bibnamefont {Amoruso}},\ }\href@noop {} {\bibfield
  {journal} {\bibinfo  {journal} {Applied Surface Science}\ }\textbf {\bibinfo
  {volume} {528}},\ \bibinfo {pages} {146607} (\bibinfo {year}
  {2020})}\BibitemShut {NoStop}%
\bibitem [{\citenamefont {Nivas}\ \emph {et~al.}(2021)\citenamefont {Nivas},
  \citenamefont {Allahyari}, \citenamefont {Skoulas}, \citenamefont {Bruzzese},
  \citenamefont {Fittipaldi}, \citenamefont {Tsibidis}, \citenamefont
  {Stratakis},\ and\ \citenamefont {Amoruso}}]{nivas2021incident}%
  \BibitemOpen
  \bibfield  {author} {\bibinfo {author} {\bibfnamefont {J.~J.}\ \bibnamefont
  {Nivas}}, \bibinfo {author} {\bibfnamefont {E.}~\bibnamefont {Allahyari}},
  \bibinfo {author} {\bibfnamefont {E.}~\bibnamefont {Skoulas}}, \bibinfo
  {author} {\bibfnamefont {R.}~\bibnamefont {Bruzzese}}, \bibinfo {author}
  {\bibfnamefont {R.}~\bibnamefont {Fittipaldi}}, \bibinfo {author}
  {\bibfnamefont {G.}~\bibnamefont {Tsibidis}}, \bibinfo {author}
  {\bibfnamefont {E.}~\bibnamefont {Stratakis}}, \ and\ \bibinfo {author}
  {\bibfnamefont {S.}~\bibnamefont {Amoruso}},\ }\href@noop {} {\bibfield
  {journal} {\bibinfo  {journal} {Applied Surface Science}\ }\textbf {\bibinfo
  {volume} {570}},\ \bibinfo {pages} {151150} (\bibinfo {year}
  {2021})}\BibitemShut {NoStop}%
\bibitem [{\citenamefont {Hu}\ \emph {et~al.}(2022)\citenamefont {Hu},
  \citenamefont {Nivas}, \citenamefont {Valadan}, \citenamefont {Fittipaldi},
  \citenamefont {Vecchione}, \citenamefont {Bruzzese}, \citenamefont
  {Altucci},\ and\ \citenamefont {Amoruso}}]{hu2022ultrafast}%
  \BibitemOpen
  \bibfield  {author} {\bibinfo {author} {\bibfnamefont {M.}~\bibnamefont
  {Hu}}, \bibinfo {author} {\bibfnamefont {J.~J.}\ \bibnamefont {Nivas}},
  \bibinfo {author} {\bibfnamefont {M.}~\bibnamefont {Valadan}}, \bibinfo
  {author} {\bibfnamefont {R.}~\bibnamefont {Fittipaldi}}, \bibinfo {author}
  {\bibfnamefont {A.}~\bibnamefont {Vecchione}}, \bibinfo {author}
  {\bibfnamefont {R.}~\bibnamefont {Bruzzese}}, \bibinfo {author}
  {\bibfnamefont {C.}~\bibnamefont {Altucci}}, \ and\ \bibinfo {author}
  {\bibfnamefont {S.}~\bibnamefont {Amoruso}},\ }\href@noop {} {\bibfield
  {journal} {\bibinfo  {journal} {Applied Surface Science}\ }\textbf {\bibinfo
  {volume} {606}},\ \bibinfo {pages} {154869} (\bibinfo {year}
  {2022})}\BibitemShut {NoStop}%
\bibitem [{\citenamefont {Kawabata}\ \emph {et~al.}(2023)\citenamefont
  {Kawabata}, \citenamefont {Bai}, \citenamefont {Obata}, \citenamefont
  {Miyaji},\ and\ \citenamefont {Sugioka}}]{kawabata2023two}%
  \BibitemOpen
  \bibfield  {author} {\bibinfo {author} {\bibfnamefont {S.}~\bibnamefont
  {Kawabata}}, \bibinfo {author} {\bibfnamefont {S.}~\bibnamefont {Bai}},
  \bibinfo {author} {\bibfnamefont {K.}~\bibnamefont {Obata}}, \bibinfo
  {author} {\bibfnamefont {G.}~\bibnamefont {Miyaji}}, \ and\ \bibinfo {author}
  {\bibfnamefont {K.}~\bibnamefont {Sugioka}},\ }\href@noop {} {\bibfield
  {journal} {\bibinfo  {journal} {International Journal of Extreme
  Manufacturing}\ } (\bibinfo {year} {2023})}\BibitemShut {NoStop}%
\bibitem [{\citenamefont {Tsibidis}\ \emph
  {et~al.}(2015{\natexlab{a}})\citenamefont {Tsibidis}, \citenamefont
  {Fotakis},\ and\ \citenamefont {Stratakis}}]{tsibidis2015ripples}%
  \BibitemOpen
  \bibfield  {author} {\bibinfo {author} {\bibfnamefont {G.~D.}\ \bibnamefont
  {Tsibidis}}, \bibinfo {author} {\bibfnamefont {C.}~\bibnamefont {Fotakis}}, \
  and\ \bibinfo {author} {\bibfnamefont {E.}~\bibnamefont {Stratakis}},\
  }\href@noop {} {\bibfield  {journal} {\bibinfo  {journal} {Physical Review
  B}\ }\textbf {\bibinfo {volume} {92}},\ \bibinfo {pages} {041405} (\bibinfo
  {year} {2015}{\natexlab{a}})}\BibitemShut {NoStop}%
\bibitem [{\citenamefont {Plateau}(1873)}]{plateau1873statique}%
  \BibitemOpen
  \bibfield  {author} {\bibinfo {author} {\bibfnamefont {J.}~\bibnamefont
  {Plateau}},\ }\href {https://books.google.de/books?id=tdf8zAEACAAJ} {\emph
  {\bibinfo {title} {Statique exp{\'e}rimentale et th{\'e}orique des liquides
  soumis aux seules forces mol{\'e}culaires}}},\ \bibinfo {series} {Statique
  exp{\'e}rimentale et th{\'e}orique des liquides soumis aux seules forces
  mol{\'e}culaires}\ No.\ \bibinfo {number} {Bd. 2,Nr. 1873}\ (\bibinfo
  {publisher} {Gauthier-Villars},\ \bibinfo {year} {1873})\BibitemShut
  {NoStop}%
\bibitem [{\citenamefont {Rayleigh}(1892)}]{rayleigh1892xvi}%
  \BibitemOpen
  \bibfield  {author} {\bibinfo {author} {\bibfnamefont {L.}~\bibnamefont
  {Rayleigh}},\ }\href {https://doi.org/10.1080/14786449208620301} {\bibfield
  {journal} {\bibinfo  {journal} {The London, Edinburgh, and Dublin
  Philosophical Magazine and Journal of Science}\ }\textbf {\bibinfo {volume}
  {34}},\ \bibinfo {pages} {145} (\bibinfo {year} {1892})}\BibitemShut
  {NoStop}%
\bibitem [{\citenamefont {Kulchin}\ \emph {et~al.}(2014)\citenamefont
  {Kulchin}, \citenamefont {Vitrik}, \citenamefont {Kuchmizhak}, \citenamefont
  {Emel'Yanov}, \citenamefont {Ionin}, \citenamefont {Kudryashov},\ and\
  \citenamefont {Makarov}}]{kulchin2014formation}%
  \BibitemOpen
  \bibfield  {author} {\bibinfo {author} {\bibfnamefont {Y.~N.}\ \bibnamefont
  {Kulchin}}, \bibinfo {author} {\bibfnamefont {O.}~\bibnamefont {Vitrik}},
  \bibinfo {author} {\bibfnamefont {A.}~\bibnamefont {Kuchmizhak}}, \bibinfo
  {author} {\bibfnamefont {V.}~\bibnamefont {Emel'Yanov}}, \bibinfo {author}
  {\bibfnamefont {A.}~\bibnamefont {Ionin}}, \bibinfo {author} {\bibfnamefont
  {S.}~\bibnamefont {Kudryashov}}, \ and\ \bibinfo {author} {\bibfnamefont
  {S.}~\bibnamefont {Makarov}},\ }\href@noop {} {\bibfield  {journal} {\bibinfo
   {journal} {Physical Review E}\ }\textbf {\bibinfo {volume} {90}},\ \bibinfo
  {pages} {023017} (\bibinfo {year} {2014})}\BibitemShut {NoStop}%
\bibitem [{\citenamefont {Pavlov}\ \emph {et~al.}(2020)\citenamefont {Pavlov},
  \citenamefont {Gurbatov}, \citenamefont {Kudryashov}, \citenamefont
  {Gurevich},\ and\ \citenamefont {Kuchmizhak}}]{pavlov2020nanocrowns}%
  \BibitemOpen
  \bibfield  {author} {\bibinfo {author} {\bibfnamefont {D.~V.}\ \bibnamefont
  {Pavlov}}, \bibinfo {author} {\bibfnamefont {S.~O.}\ \bibnamefont
  {Gurbatov}}, \bibinfo {author} {\bibfnamefont {S.~I.}\ \bibnamefont
  {Kudryashov}}, \bibinfo {author} {\bibfnamefont {E.~L.}\ \bibnamefont
  {Gurevich}}, \ and\ \bibinfo {author} {\bibfnamefont {A.~A.}\ \bibnamefont
  {Kuchmizhak}},\ }\href@noop {} {\bibfield  {journal} {\bibinfo  {journal}
  {Applied Surface Science}\ }\textbf {\bibinfo {volume} {511}},\ \bibinfo
  {pages} {145463} (\bibinfo {year} {2020})}\BibitemShut {NoStop}%
\bibitem [{\citenamefont {Fujii}\ \emph {et~al.}(1999)\citenamefont {Fujii},
  \citenamefont {Yamamoto}, \citenamefont {Hara},\ and\ \citenamefont
  {Nogi}}]{SiCA}%
  \BibitemOpen
  \bibfield  {author} {\bibinfo {author} {\bibfnamefont {H.}~\bibnamefont
  {Fujii}}, \bibinfo {author} {\bibfnamefont {M.}~\bibnamefont {Yamamoto}},
  \bibinfo {author} {\bibfnamefont {S.}~\bibnamefont {Hara}}, \ and\ \bibinfo
  {author} {\bibfnamefont {K.}~\bibnamefont {Nogi}},\ }\href {\doibase
  10.1023/A:1004673605025} {\bibfield  {journal} {\bibinfo  {journal} {J.
  Mater. Sci.}\ }\textbf {\bibinfo {volume} {34}},\ \bibinfo {pages} {3165}
  (\bibinfo {year} {1999})}\BibitemShut {NoStop}%
\bibitem [{\citenamefont {Eustathopoulos}\ and\ \citenamefont
  {Drevet}(2013)}]{EUSTATHOPOULOS201377}%
  \BibitemOpen
  \bibfield  {author} {\bibinfo {author} {\bibfnamefont {N.}~\bibnamefont
  {Eustathopoulos}}\ and\ \bibinfo {author} {\bibfnamefont {B.}~\bibnamefont
  {Drevet}},\ }\href {\doibase https://doi.org/10.1016/j.jcrysgro.2013.02.010}
  {\bibfield  {journal} {\bibinfo  {journal} {Journal of Crystal Growth}\
  }\textbf {\bibinfo {volume} {371}},\ \bibinfo {pages} {77} (\bibinfo {year}
  {2013})}\BibitemShut {NoStop}%
\bibitem [{\citenamefont {Gurevich}(2016)}]{gurevich2016mechanisms}%
  \BibitemOpen
  \bibfield  {author} {\bibinfo {author} {\bibfnamefont {E.~L.}\ \bibnamefont
  {Gurevich}},\ }\href@noop {} {\bibfield  {journal} {\bibinfo  {journal}
  {Applied Surface Science}\ }\textbf {\bibinfo {volume} {374}},\ \bibinfo
  {pages} {56} (\bibinfo {year} {2016})}\BibitemShut {NoStop}%
\bibitem [{\citenamefont {Sasaki}\ \emph {et~al.}(1995)\citenamefont {Sasaki},
  \citenamefont {Tokizaki}, \citenamefont {Huang}, \citenamefont {Terashima},\
  and\ \citenamefont {Kimura}}]{Sasaki1995viscosity}%
  \BibitemOpen
  \bibfield  {author} {\bibinfo {author} {\bibfnamefont {H.}~\bibnamefont
  {Sasaki}}, \bibinfo {author} {\bibfnamefont {E.}~\bibnamefont {Tokizaki}},
  \bibinfo {author} {\bibfnamefont {X.~M.}\ \bibnamefont {Huang}}, \bibinfo
  {author} {\bibfnamefont {K.}~\bibnamefont {Terashima}}, \ and\ \bibinfo
  {author} {\bibfnamefont {S.~K.~S.}\ \bibnamefont {Kimura}},\ }\href {\doibase
  10.1143/JJAP.34.3432} {\bibfield  {journal} {\bibinfo  {journal} {Japanese
  Journal of Applied Physics}\ }\textbf {\bibinfo {volume} {34}},\ \bibinfo
  {pages} {3432} (\bibinfo {year} {1995})}\BibitemShut {NoStop}%
\bibitem [{\citenamefont {Gundrum}\ \emph {et~al.}(2007)\citenamefont
  {Gundrum}, \citenamefont {Averback},\ and\ \citenamefont
  {Cahill}}]{Gundrum2007Sisolodification}%
  \BibitemOpen
  \bibfield  {author} {\bibinfo {author} {\bibfnamefont {B.~C.}\ \bibnamefont
  {Gundrum}}, \bibinfo {author} {\bibfnamefont {R.~S.}\ \bibnamefont
  {Averback}}, \ and\ \bibinfo {author} {\bibfnamefont {D.~G.}\ \bibnamefont
  {Cahill}},\ }\href {\doibase 10.1063/1.2752731} {\bibfield  {journal}
  {\bibinfo  {journal} {Appl. Phys. Lett.}\ }\textbf {\bibinfo {volume} {91}}
  (\bibinfo {year} {2007}),\ 10.1063/1.2752731}\BibitemShut {NoStop}%
\bibitem [{\citenamefont {Yamamoto}\ \emph {et~al.}(1991)\citenamefont
  {Yamamoto}, \citenamefont {Abe},\ and\ \citenamefont {ichiro Takasu
  Shin-ichiro Takasu}}]{Yamamoto1991}%
  \BibitemOpen
  \bibfield  {author} {\bibinfo {author} {\bibfnamefont {K.}~\bibnamefont
  {Yamamoto}}, \bibinfo {author} {\bibfnamefont {T.}~\bibnamefont {Abe}}, \
  and\ \bibinfo {author} {\bibfnamefont {S.}~\bibnamefont {ichiro Takasu
  Shin-ichiro Takasu}},\ }\href {\doibase 10.1143/JJAP.30.2423} {\bibfield
  {journal} {\bibinfo  {journal} {Japanese Journal of Applied Physics}\
  }\textbf {\bibinfo {volume} {30}},\ \bibinfo {pages} {2423} (\bibinfo {year}
  {1991})}\BibitemShut {NoStop}%
\bibitem [{\citenamefont {Otto}\ \emph {et~al.}(2006)\citenamefont {Otto},
  \citenamefont {Rump},\ and\ \citenamefont {Slepcev}}]{OttoDrops}%
  \BibitemOpen
  \bibfield  {author} {\bibinfo {author} {\bibfnamefont {F.}~\bibnamefont
  {Otto}}, \bibinfo {author} {\bibfnamefont {T.}~\bibnamefont {Rump}}, \ and\
  \bibinfo {author} {\bibfnamefont {D.}~\bibnamefont {Slepcev}},\ }\href
  {\doibase 10.1137/050630192} {\bibfield  {journal} {\bibinfo  {journal} {SIAM
  Journal on Mathematical Analysis}\ }\textbf {\bibinfo {volume} {38}},\
  \bibinfo {pages} {503} (\bibinfo {year} {2006})},\ \Eprint
  {http://arxiv.org/abs/https://doi.org/10.1137/050630192}
  {https://doi.org/10.1137/050630192} \BibitemShut {NoStop}%
\bibitem [{\citenamefont {Liu}\ \emph {et~al.}(2021)\citenamefont {Liu},
  \citenamefont {Hu}, \citenamefont {Jiang}, \citenamefont {Huang},
  \citenamefont {Lu}, \citenamefont {Yin}, \citenamefont {Qiu}, \citenamefont
  {Liu}, \citenamefont {Li}, \citenamefont {Wang} \emph
  {et~al.}}]{liu2021formation}%
  \BibitemOpen
  \bibfield  {author} {\bibinfo {author} {\bibfnamefont {W.}~\bibnamefont
  {Liu}}, \bibinfo {author} {\bibfnamefont {J.}~\bibnamefont {Hu}}, \bibinfo
  {author} {\bibfnamefont {L.}~\bibnamefont {Jiang}}, \bibinfo {author}
  {\bibfnamefont {J.}~\bibnamefont {Huang}}, \bibinfo {author} {\bibfnamefont
  {J.}~\bibnamefont {Lu}}, \bibinfo {author} {\bibfnamefont {J.}~\bibnamefont
  {Yin}}, \bibinfo {author} {\bibfnamefont {Z.}~\bibnamefont {Qiu}}, \bibinfo
  {author} {\bibfnamefont {H.}~\bibnamefont {Liu}}, \bibinfo {author}
  {\bibfnamefont {C.}~\bibnamefont {Li}}, \bibinfo {author} {\bibfnamefont
  {S.}~\bibnamefont {Wang}},  \emph {et~al.},\ }\href@noop {} {\bibfield
  {journal} {\bibinfo  {journal} {Nanophotonics}\ }\textbf {\bibinfo {volume}
  {10}},\ \bibinfo {pages} {1273} (\bibinfo {year} {2021})}\BibitemShut
  {NoStop}%
\bibitem [{\citenamefont {Aspnes}\ and\ \citenamefont
  {Studna}(1983)}]{aspnes1983dielectric}%
  \BibitemOpen
  \bibfield  {author} {\bibinfo {author} {\bibfnamefont {D.~E.}\ \bibnamefont
  {Aspnes}}\ and\ \bibinfo {author} {\bibfnamefont {A.}~\bibnamefont
  {Studna}},\ }\href@noop {} {\bibfield  {journal} {\bibinfo  {journal}
  {Physical review B}\ }\textbf {\bibinfo {volume} {27}},\ \bibinfo {pages}
  {985} (\bibinfo {year} {1983})}\BibitemShut {NoStop}%
\bibitem [{\citenamefont {Liu}\ \emph {et~al.}(2023)\citenamefont {Liu},
  \citenamefont {Ding}, \citenamefont {Xie}, \citenamefont {Xu}, \citenamefont
  {Jeong},\ and\ \citenamefont {Yang}}]{liu2023one}%
  \BibitemOpen
  \bibfield  {author} {\bibinfo {author} {\bibfnamefont {Y.}~\bibnamefont
  {Liu}}, \bibinfo {author} {\bibfnamefont {Y.}~\bibnamefont {Ding}}, \bibinfo
  {author} {\bibfnamefont {J.}~\bibnamefont {Xie}}, \bibinfo {author}
  {\bibfnamefont {L.}~\bibnamefont {Xu}}, \bibinfo {author} {\bibfnamefont
  {I.~W.}\ \bibnamefont {Jeong}}, \ and\ \bibinfo {author} {\bibfnamefont
  {L.}~\bibnamefont {Yang}},\ }\href@noop {} {\bibfield  {journal} {\bibinfo
  {journal} {Materials \& Design}\ }\textbf {\bibinfo {volume} {225}},\
  \bibinfo {pages} {111443} (\bibinfo {year} {2023})}\BibitemShut {NoStop}%
\bibitem [{\citenamefont {Nivas}\ \emph {et~al.}(2015)\citenamefont {Nivas},
  \citenamefont {Shutong}, \citenamefont {Anoop}, \citenamefont {Rubano},
  \citenamefont {Fittipaldi}, \citenamefont {Vecchione}, \citenamefont
  {Paparo}, \citenamefont {Marrucci}, \citenamefont {Bruzzese},\ and\
  \citenamefont {Amoruso}}]{nivas2015laser}%
  \BibitemOpen
  \bibfield  {author} {\bibinfo {author} {\bibfnamefont {J.~J.}\ \bibnamefont
  {Nivas}}, \bibinfo {author} {\bibfnamefont {H.}~\bibnamefont {Shutong}},
  \bibinfo {author} {\bibfnamefont {K.}~\bibnamefont {Anoop}}, \bibinfo
  {author} {\bibfnamefont {A.}~\bibnamefont {Rubano}}, \bibinfo {author}
  {\bibfnamefont {R.}~\bibnamefont {Fittipaldi}}, \bibinfo {author}
  {\bibfnamefont {A.}~\bibnamefont {Vecchione}}, \bibinfo {author}
  {\bibfnamefont {D.}~\bibnamefont {Paparo}}, \bibinfo {author} {\bibfnamefont
  {L.}~\bibnamefont {Marrucci}}, \bibinfo {author} {\bibfnamefont
  {R.}~\bibnamefont {Bruzzese}}, \ and\ \bibinfo {author} {\bibfnamefont
  {S.}~\bibnamefont {Amoruso}},\ }\href@noop {} {\bibfield  {journal} {\bibinfo
   {journal} {Optics letters}\ }\textbf {\bibinfo {volume} {40}},\ \bibinfo
  {pages} {4611} (\bibinfo {year} {2015})}\BibitemShut {NoStop}%
\bibitem [{\citenamefont {Tsibidis}\ \emph
  {et~al.}(2015{\natexlab{b}})\citenamefont {Tsibidis}, \citenamefont
  {Skoulas},\ and\ \citenamefont {Stratakis}}]{tsibidis2015ripple}%
  \BibitemOpen
  \bibfield  {author} {\bibinfo {author} {\bibfnamefont {G.~D.}\ \bibnamefont
  {Tsibidis}}, \bibinfo {author} {\bibfnamefont {E.}~\bibnamefont {Skoulas}}, \
  and\ \bibinfo {author} {\bibfnamefont {E.}~\bibnamefont {Stratakis}},\
  }\href@noop {} {\bibfield  {journal} {\bibinfo  {journal} {Optics letters}\
  }\textbf {\bibinfo {volume} {40}},\ \bibinfo {pages} {5172} (\bibinfo {year}
  {2015}{\natexlab{b}})}\BibitemShut {NoStop}%
\bibitem [{\citenamefont {Wovchko}\ \emph {et~al.}(1995)\citenamefont
  {Wovchko}, \citenamefont {Camp}, \citenamefont {Glass},\ and\ \citenamefont
  {Yates}}]{SiOCH3}%
  \BibitemOpen
  \bibfield  {author} {\bibinfo {author} {\bibfnamefont {E.~A.}\ \bibnamefont
  {Wovchko}}, \bibinfo {author} {\bibfnamefont {J.~C.}\ \bibnamefont {Camp}},
  \bibinfo {author} {\bibfnamefont {J.~A.~J.}\ \bibnamefont {Glass}}, \ and\
  \bibinfo {author} {\bibfnamefont {J.~T.}\ \bibnamefont {Yates}},\ }\href
  {\doibase 10.1021/la00007a044} {\bibfield  {journal} {\bibinfo  {journal}
  {Langmuir}\ }\textbf {\bibinfo {volume} {11}},\ \bibinfo {pages} {2592}
  (\bibinfo {year} {1995})},\ \Eprint
  {http://arxiv.org/abs/https://doi.org/10.1021/la00007a044}
  {https://doi.org/10.1021/la00007a044} \BibitemShut {NoStop}%
\bibitem [{\citenamefont {Diebold}\ \emph {et~al.}(2009)\citenamefont
  {Diebold}, \citenamefont {Mack}, \citenamefont {Doorn},\ and\ \citenamefont
  {Mazur}}]{diebold2009femtosecond}%
  \BibitemOpen
  \bibfield  {author} {\bibinfo {author} {\bibfnamefont {E.~D.}\ \bibnamefont
  {Diebold}}, \bibinfo {author} {\bibfnamefont {N.~H.}\ \bibnamefont {Mack}},
  \bibinfo {author} {\bibfnamefont {S.~K.}\ \bibnamefont {Doorn}}, \ and\
  \bibinfo {author} {\bibfnamefont {E.}~\bibnamefont {Mazur}},\ }\href@noop {}
  {\bibfield  {journal} {\bibinfo  {journal} {Langmuir}\ }\textbf {\bibinfo
  {volume} {25}},\ \bibinfo {pages} {1790} (\bibinfo {year}
  {2009})}\BibitemShut {NoStop}%
\bibitem [{\citenamefont {Erk{\i}zan}\ \emph {et~al.}(2022)\citenamefont
  {Erk{\i}zan}, \citenamefont {{\.I}dikut}, \citenamefont {Demirta{\c{s}}},
  \citenamefont {Goodarzi}, \citenamefont {Demir}, \citenamefont {Borra},
  \citenamefont {Pavlov},\ and\ \citenamefont {Bek}}]{erkizan2022lipss}%
  \BibitemOpen
  \bibfield  {author} {\bibinfo {author} {\bibfnamefont {S.~N.}\ \bibnamefont
  {Erk{\i}zan}}, \bibinfo {author} {\bibfnamefont {F.}~\bibnamefont
  {{\.I}dikut}}, \bibinfo {author} {\bibfnamefont {{\"O}.}~\bibnamefont
  {Demirta{\c{s}}}}, \bibinfo {author} {\bibfnamefont {A.}~\bibnamefont
  {Goodarzi}}, \bibinfo {author} {\bibfnamefont {A.~K.}\ \bibnamefont {Demir}},
  \bibinfo {author} {\bibfnamefont {M.}~\bibnamefont {Borra}}, \bibinfo
  {author} {\bibfnamefont {I.}~\bibnamefont {Pavlov}}, \ and\ \bibinfo {author}
  {\bibfnamefont {A.}~\bibnamefont {Bek}},\ }\href@noop {} {\bibfield
  {journal} {\bibinfo  {journal} {Advanced Optical Materials}\ }\textbf
  {\bibinfo {volume} {10}},\ \bibinfo {pages} {2200233} (\bibinfo {year}
  {2022})}\BibitemShut {NoStop}%
\bibitem [{\citenamefont {Alessandri}\ and\ \citenamefont
  {Lombardi}(2016)}]{alessandri2016enhanced}%
  \BibitemOpen
  \bibfield  {author} {\bibinfo {author} {\bibfnamefont {I.}~\bibnamefont
  {Alessandri}}\ and\ \bibinfo {author} {\bibfnamefont {J.~R.}\ \bibnamefont
  {Lombardi}},\ }\href@noop {} {\bibfield  {journal} {\bibinfo  {journal}
  {Chemical reviews}\ }\textbf {\bibinfo {volume} {116}},\ \bibinfo {pages}
  {14921} (\bibinfo {year} {2016})}\BibitemShut {NoStop}%
\bibitem [{\citenamefont {Tittl}\ \emph {et~al.}(2018)\citenamefont {Tittl},
  \citenamefont {Leitis}, \citenamefont {Liu}, \citenamefont {Yesilkoy},
  \citenamefont {Choi}, \citenamefont {Neshev}, \citenamefont {Kivshar},\ and\
  \citenamefont {Altug}}]{tittl2018imaging}%
  \BibitemOpen
  \bibfield  {author} {\bibinfo {author} {\bibfnamefont {A.}~\bibnamefont
  {Tittl}}, \bibinfo {author} {\bibfnamefont {A.}~\bibnamefont {Leitis}},
  \bibinfo {author} {\bibfnamefont {M.}~\bibnamefont {Liu}}, \bibinfo {author}
  {\bibfnamefont {F.}~\bibnamefont {Yesilkoy}}, \bibinfo {author}
  {\bibfnamefont {D.-Y.}\ \bibnamefont {Choi}}, \bibinfo {author}
  {\bibfnamefont {D.~N.}\ \bibnamefont {Neshev}}, \bibinfo {author}
  {\bibfnamefont {Y.~S.}\ \bibnamefont {Kivshar}}, \ and\ \bibinfo {author}
  {\bibfnamefont {H.}~\bibnamefont {Altug}},\ }\href@noop {} {\bibfield
  {journal} {\bibinfo  {journal} {Science}\ }\textbf {\bibinfo {volume}
  {360}},\ \bibinfo {pages} {1105} (\bibinfo {year} {2018})}\BibitemShut
  {NoStop}%
\bibitem [{\citenamefont {Mitsai}\ \emph {et~al.}(2018)\citenamefont {Mitsai},
  \citenamefont {Kuchmizhak}, \citenamefont {Pustovalov}, \citenamefont
  {Sergeev}, \citenamefont {Mironenko}, \citenamefont {Bratskaya},
  \citenamefont {Linklater}, \citenamefont {Bal{\v{c}}ytis}, \citenamefont
  {Ivanova},\ and\ \citenamefont {Juodkazis}}]{mitsai2018chemically}%
  \BibitemOpen
  \bibfield  {author} {\bibinfo {author} {\bibfnamefont {E.}~\bibnamefont
  {Mitsai}}, \bibinfo {author} {\bibfnamefont {A.}~\bibnamefont {Kuchmizhak}},
  \bibinfo {author} {\bibfnamefont {E.}~\bibnamefont {Pustovalov}}, \bibinfo
  {author} {\bibfnamefont {A.}~\bibnamefont {Sergeev}}, \bibinfo {author}
  {\bibfnamefont {A.}~\bibnamefont {Mironenko}}, \bibinfo {author}
  {\bibfnamefont {S.}~\bibnamefont {Bratskaya}}, \bibinfo {author}
  {\bibfnamefont {D.}~\bibnamefont {Linklater}}, \bibinfo {author}
  {\bibfnamefont {A.}~\bibnamefont {Bal{\v{c}}ytis}}, \bibinfo {author}
  {\bibfnamefont {E.}~\bibnamefont {Ivanova}}, \ and\ \bibinfo {author}
  {\bibfnamefont {S.}~\bibnamefont {Juodkazis}},\ }\href@noop {} {\bibfield
  {journal} {\bibinfo  {journal} {Nanoscale}\ }\textbf {\bibinfo {volume}
  {10}},\ \bibinfo {pages} {9780} (\bibinfo {year} {2018})}\BibitemShut
  {NoStop}%
\bibitem [{\citenamefont {Mironenko}\ \emph {et~al.}(2019)\citenamefont
  {Mironenko}, \citenamefont {Tutov}, \citenamefont {Sergeev}, \citenamefont
  {Mitsai}, \citenamefont {Ustinov}, \citenamefont {Zhizhchenko}, \citenamefont
  {Linklater}, \citenamefont {Bratskaya}, \citenamefont {Juodkazis},\ and\
  \citenamefont {Kuchmizhak}}]{mironenko2019ultratrace}%
  \BibitemOpen
  \bibfield  {author} {\bibinfo {author} {\bibfnamefont {A.~Y.}\ \bibnamefont
  {Mironenko}}, \bibinfo {author} {\bibfnamefont {M.~V.}\ \bibnamefont
  {Tutov}}, \bibinfo {author} {\bibfnamefont {A.~A.}\ \bibnamefont {Sergeev}},
  \bibinfo {author} {\bibfnamefont {E.~V.}\ \bibnamefont {Mitsai}}, \bibinfo
  {author} {\bibfnamefont {A.~Y.}\ \bibnamefont {Ustinov}}, \bibinfo {author}
  {\bibfnamefont {A.~Y.}\ \bibnamefont {Zhizhchenko}}, \bibinfo {author}
  {\bibfnamefont {D.~P.}\ \bibnamefont {Linklater}}, \bibinfo {author}
  {\bibfnamefont {S.~Y.}\ \bibnamefont {Bratskaya}}, \bibinfo {author}
  {\bibfnamefont {S.}~\bibnamefont {Juodkazis}}, \ and\ \bibinfo {author}
  {\bibfnamefont {A.~A.}\ \bibnamefont {Kuchmizhak}},\ }\href@noop {}
  {\bibfield  {journal} {\bibinfo  {journal} {ACS sensors}\ }\textbf {\bibinfo
  {volume} {4}},\ \bibinfo {pages} {2879} (\bibinfo {year} {2019})}\BibitemShut
  {NoStop}%
\bibitem [{\citenamefont {Dostovalov}\ \emph {et~al.}(2020)\citenamefont
  {Dostovalov}, \citenamefont {Bronnikov}, \citenamefont {Korolkov},
  \citenamefont {Babin}, \citenamefont {Mitsai}, \citenamefont {Mironenko},
  \citenamefont {Tutov}, \citenamefont {Zhang}, \citenamefont {Sugioka},
  \citenamefont {Maksimovic} \emph {et~al.}}]{dostovalov2020hierarchical}%
  \BibitemOpen
  \bibfield  {author} {\bibinfo {author} {\bibfnamefont {A.}~\bibnamefont
  {Dostovalov}}, \bibinfo {author} {\bibfnamefont {K.}~\bibnamefont
  {Bronnikov}}, \bibinfo {author} {\bibfnamefont {V.}~\bibnamefont {Korolkov}},
  \bibinfo {author} {\bibfnamefont {S.}~\bibnamefont {Babin}}, \bibinfo
  {author} {\bibfnamefont {E.}~\bibnamefont {Mitsai}}, \bibinfo {author}
  {\bibfnamefont {A.}~\bibnamefont {Mironenko}}, \bibinfo {author}
  {\bibfnamefont {M.}~\bibnamefont {Tutov}}, \bibinfo {author} {\bibfnamefont
  {D.}~\bibnamefont {Zhang}}, \bibinfo {author} {\bibfnamefont
  {K.}~\bibnamefont {Sugioka}}, \bibinfo {author} {\bibfnamefont
  {J.}~\bibnamefont {Maksimovic}},  \emph {et~al.},\ }\href@noop {} {\bibfield
  {journal} {\bibinfo  {journal} {Nanoscale}\ }\textbf {\bibinfo {volume}
  {12}},\ \bibinfo {pages} {13431} (\bibinfo {year} {2020})}\BibitemShut
  {NoStop}%
\bibitem [{\citenamefont {Sher}\ \emph {et~al.}(2011)\citenamefont {Sher},
  \citenamefont {Winkler},\ and\ \citenamefont {Mazur}}]{sher2011pulsed}%
  \BibitemOpen
  \bibfield  {author} {\bibinfo {author} {\bibfnamefont {M.-J.}\ \bibnamefont
  {Sher}}, \bibinfo {author} {\bibfnamefont {M.~T.}\ \bibnamefont {Winkler}}, \
  and\ \bibinfo {author} {\bibfnamefont {E.}~\bibnamefont {Mazur}},\
  }\href@noop {} {\bibfield  {journal} {\bibinfo  {journal} {MRS bulletin}\
  }\textbf {\bibinfo {volume} {36}},\ \bibinfo {pages} {439} (\bibinfo {year}
  {2011})}\BibitemShut {NoStop}%
\bibitem [{\citenamefont {Zielke}\ \emph {et~al.}(2012)\citenamefont {Zielke},
  \citenamefont {Sylla}, \citenamefont {Neubert}, \citenamefont {Brendel},\
  and\ \citenamefont {Schmidt}}]{zielke2012direct}%
  \BibitemOpen
  \bibfield  {author} {\bibinfo {author} {\bibfnamefont {D.}~\bibnamefont
  {Zielke}}, \bibinfo {author} {\bibfnamefont {D.}~\bibnamefont {Sylla}},
  \bibinfo {author} {\bibfnamefont {T.}~\bibnamefont {Neubert}}, \bibinfo
  {author} {\bibfnamefont {R.}~\bibnamefont {Brendel}}, \ and\ \bibinfo
  {author} {\bibfnamefont {J.}~\bibnamefont {Schmidt}},\ }\href@noop {}
  {\bibfield  {journal} {\bibinfo  {journal} {IEEE Journal of Photovoltaics}\
  }\textbf {\bibinfo {volume} {3}},\ \bibinfo {pages} {656} (\bibinfo {year}
  {2012})}\BibitemShut {NoStop}%
\bibitem [{\citenamefont {Juntunen}\ \emph {et~al.}(2016)\citenamefont
  {Juntunen}, \citenamefont {Heinonen}, \citenamefont {V{\"a}h{\"a}nissi},
  \citenamefont {Repo}, \citenamefont {Valluru},\ and\ \citenamefont
  {Savin}}]{juntunen2016near}%
  \BibitemOpen
  \bibfield  {author} {\bibinfo {author} {\bibfnamefont {M.~A.}\ \bibnamefont
  {Juntunen}}, \bibinfo {author} {\bibfnamefont {J.}~\bibnamefont {Heinonen}},
  \bibinfo {author} {\bibfnamefont {V.}~\bibnamefont {V{\"a}h{\"a}nissi}},
  \bibinfo {author} {\bibfnamefont {P.}~\bibnamefont {Repo}}, \bibinfo {author}
  {\bibfnamefont {D.}~\bibnamefont {Valluru}}, \ and\ \bibinfo {author}
  {\bibfnamefont {H.}~\bibnamefont {Savin}},\ }\href@noop {} {\bibfield
  {journal} {\bibinfo  {journal} {Nature Photonics}\ }\textbf {\bibinfo
  {volume} {10}},\ \bibinfo {pages} {777} (\bibinfo {year} {2016})}\BibitemShut
  {NoStop}%
\bibitem [{\citenamefont {Garin}\ \emph {et~al.}(2020)\citenamefont {Garin},
  \citenamefont {Heinonen}, \citenamefont {Werner}, \citenamefont {Pasanen},
  \citenamefont {V{\"a}h{\"a}nissi}, \citenamefont {Haarahiltunen},
  \citenamefont {Juntunen},\ and\ \citenamefont {Savin}}]{garin2020black}%
  \BibitemOpen
  \bibfield  {author} {\bibinfo {author} {\bibfnamefont {M.}~\bibnamefont
  {Garin}}, \bibinfo {author} {\bibfnamefont {J.}~\bibnamefont {Heinonen}},
  \bibinfo {author} {\bibfnamefont {L.}~\bibnamefont {Werner}}, \bibinfo
  {author} {\bibfnamefont {T.}~\bibnamefont {Pasanen}}, \bibinfo {author}
  {\bibfnamefont {V.}~\bibnamefont {V{\"a}h{\"a}nissi}}, \bibinfo {author}
  {\bibfnamefont {A.}~\bibnamefont {Haarahiltunen}}, \bibinfo {author}
  {\bibfnamefont {M.}~\bibnamefont {Juntunen}}, \ and\ \bibinfo {author}
  {\bibfnamefont {H.}~\bibnamefont {Savin}},\ }\href@noop {} {\bibfield
  {journal} {\bibinfo  {journal} {Physical Review Letters}\ }\textbf {\bibinfo
  {volume} {125}},\ \bibinfo {pages} {117702} (\bibinfo {year}
  {2020})}\BibitemShut {NoStop}%
\bibitem [{\citenamefont {Seta?la?}\ \emph {et~al.}(2023)\citenamefont
  {Seta?la?}, \citenamefont {Chen}, \citenamefont {Pasanen}, \citenamefont
  {Liu}, \citenamefont {Radfar}, \citenamefont {Va?ha?nissi},\ and\
  \citenamefont {Savin}}]{seta?la?2023boron}%
  \BibitemOpen
  \bibfield  {author} {\bibinfo {author} {\bibfnamefont {O.~E.}\ \bibnamefont
  {Seta?la?}}, \bibinfo {author} {\bibfnamefont {K.}~\bibnamefont {Chen}},
  \bibinfo {author} {\bibfnamefont {T.~P.}\ \bibnamefont {Pasanen}}, \bibinfo
  {author} {\bibfnamefont {X.}~\bibnamefont {Liu}}, \bibinfo {author}
  {\bibfnamefont {B.}~\bibnamefont {Radfar}}, \bibinfo {author} {\bibfnamefont
  {V.}~\bibnamefont {Va?ha?nissi}}, \ and\ \bibinfo {author} {\bibfnamefont
  {H.}~\bibnamefont {Savin}},\ }\href@noop {} {\bibfield  {journal} {\bibinfo
  {journal} {ACS photonics}\ } (\bibinfo {year} {2023})}\BibitemShut {NoStop}%
\bibitem [{\citenamefont {Zhao}\ \emph {et~al.}(2023)\citenamefont {Zhao},
  \citenamefont {Zhang}, \citenamefont {Jing}, \citenamefont {Gao},
  \citenamefont {Liao}, \citenamefont {Zhang}, \citenamefont {Liu},
  \citenamefont {Wang}, \citenamefont {Wang},\ and\ \citenamefont
  {Xue}}]{zhao2023black}%
  \BibitemOpen
  \bibfield  {author} {\bibinfo {author} {\bibfnamefont {Z.}~\bibnamefont
  {Zhao}}, \bibinfo {author} {\bibfnamefont {Z.}~\bibnamefont {Zhang}},
  \bibinfo {author} {\bibfnamefont {J.}~\bibnamefont {Jing}}, \bibinfo {author}
  {\bibfnamefont {R.}~\bibnamefont {Gao}}, \bibinfo {author} {\bibfnamefont
  {Z.}~\bibnamefont {Liao}}, \bibinfo {author} {\bibfnamefont {W.}~\bibnamefont
  {Zhang}}, \bibinfo {author} {\bibfnamefont {G.}~\bibnamefont {Liu}}, \bibinfo
  {author} {\bibfnamefont {Y.}~\bibnamefont {Wang}}, \bibinfo {author}
  {\bibfnamefont {K.}~\bibnamefont {Wang}}, \ and\ \bibinfo {author}
  {\bibfnamefont {C.}~\bibnamefont {Xue}},\ }\href@noop {} {\bibfield
  {journal} {\bibinfo  {journal} {APL Materials}\ }\textbf {\bibinfo {volume}
  {11}} (\bibinfo {year} {2023})}\BibitemShut {NoStop}%
\bibitem [{\citenamefont {Zhao}\ \emph {et~al.}(2020)\citenamefont {Zhao},
  \citenamefont {Li}, \citenamefont {Chen}, \citenamefont {Chen},\ and\
  \citenamefont {Sun}}]{zhao2020ultrafast}%
  \BibitemOpen
  \bibfield  {author} {\bibinfo {author} {\bibfnamefont {J.-H.}\ \bibnamefont
  {Zhao}}, \bibinfo {author} {\bibfnamefont {X.-B.}\ \bibnamefont {Li}},
  \bibinfo {author} {\bibfnamefont {Q.-D.}\ \bibnamefont {Chen}}, \bibinfo
  {author} {\bibfnamefont {Z.-G.}\ \bibnamefont {Chen}}, \ and\ \bibinfo
  {author} {\bibfnamefont {H.-B.}\ \bibnamefont {Sun}},\ }\href@noop {}
  {\bibfield  {journal} {\bibinfo  {journal} {Materials Today Nano}\ }\textbf
  {\bibinfo {volume} {11}},\ \bibinfo {pages} {100078} (\bibinfo {year}
  {2020})}\BibitemShut {NoStop}%
\bibitem [{\citenamefont {Li}\ \emph {et~al.}(2018)\citenamefont {Li},
  \citenamefont {Zhao}, \citenamefont {Chen}, \citenamefont {Feng},\ and\
  \citenamefont {Sun}}]{li2018sub}%
  \BibitemOpen
  \bibfield  {author} {\bibinfo {author} {\bibfnamefont {C.-H.}\ \bibnamefont
  {Li}}, \bibinfo {author} {\bibfnamefont {J.-H.}\ \bibnamefont {Zhao}},
  \bibinfo {author} {\bibfnamefont {Q.-D.}\ \bibnamefont {Chen}}, \bibinfo
  {author} {\bibfnamefont {J.}~\bibnamefont {Feng}}, \ and\ \bibinfo {author}
  {\bibfnamefont {H.-B.}\ \bibnamefont {Sun}},\ }\href@noop {} {\bibfield
  {journal} {\bibinfo  {journal} {Optics Letters}\ }\textbf {\bibinfo {volume}
  {43}},\ \bibinfo {pages} {1710} (\bibinfo {year} {2018})}\BibitemShut
  {NoStop}%
\bibitem [{\citenamefont {Beresna}\ \emph {et~al.}(2011)\citenamefont
  {Beresna}, \citenamefont {Gecevi{\v{c}}ius}, \citenamefont {Kazansky},\ and\
  \citenamefont {Gertus}}]{beresna2011radially}%
  \BibitemOpen
  \bibfield  {author} {\bibinfo {author} {\bibfnamefont {M.}~\bibnamefont
  {Beresna}}, \bibinfo {author} {\bibfnamefont {M.}~\bibnamefont
  {Gecevi{\v{c}}ius}}, \bibinfo {author} {\bibfnamefont {P.~G.}\ \bibnamefont
  {Kazansky}}, \ and\ \bibinfo {author} {\bibfnamefont {T.}~\bibnamefont
  {Gertus}},\ }\href@noop {} {\bibfield  {journal} {\bibinfo  {journal}
  {Applied Physics Letters}\ }\textbf {\bibinfo {volume} {98}},\ \bibinfo
  {pages} {201101} (\bibinfo {year} {2011})}\BibitemShut {NoStop}%
\bibitem [{\citenamefont {Liu}(1982)}]{liu1982simple}%
  \BibitemOpen
  \bibfield  {author} {\bibinfo {author} {\bibfnamefont {J.-M.}\ \bibnamefont
  {Liu}},\ }\href@noop {} {\bibfield  {journal} {\bibinfo  {journal} {Optics
  letters}\ }\textbf {\bibinfo {volume} {7}},\ \bibinfo {pages} {196} (\bibinfo
  {year} {1982})}\BibitemShut {NoStop}%
\bibitem [{\citenamefont {Zhang}\ \emph {et~al.}(2017)\citenamefont {Zhang},
  \citenamefont {Huang}, \citenamefont {Zhu}, \citenamefont {Shao},\ and\
  \citenamefont {Chen}}]{zhang2017dimensional}%
  \BibitemOpen
  \bibfield  {author} {\bibinfo {author} {\bibfnamefont {J.}~\bibnamefont
  {Zhang}}, \bibinfo {author} {\bibfnamefont {S.-J.}\ \bibnamefont {Huang}},
  \bibinfo {author} {\bibfnamefont {F.-Q.}\ \bibnamefont {Zhu}}, \bibinfo
  {author} {\bibfnamefont {W.}~\bibnamefont {Shao}}, \ and\ \bibinfo {author}
  {\bibfnamefont {M.-S.}\ \bibnamefont {Chen}},\ }\href@noop {} {\bibfield
  {journal} {\bibinfo  {journal} {Applied Optics}\ }\textbf {\bibinfo {volume}
  {56}},\ \bibinfo {pages} {3556} (\bibinfo {year} {2017})}\BibitemShut
  {NoStop}%
\bibitem [{\citenamefont {Moutzouris}\ \emph {et~al.}(2014)\citenamefont
  {Moutzouris}, \citenamefont {Papamichael}, \citenamefont {Betsis},
  \citenamefont {Stavrakas}, \citenamefont {Hloupis},\ and\ \citenamefont
  {Triantis}}]{moutzouris2014refractive}%
  \BibitemOpen
  \bibfield  {author} {\bibinfo {author} {\bibfnamefont {K.}~\bibnamefont
  {Moutzouris}}, \bibinfo {author} {\bibfnamefont {M.}~\bibnamefont
  {Papamichael}}, \bibinfo {author} {\bibfnamefont {S.~C.}\ \bibnamefont
  {Betsis}}, \bibinfo {author} {\bibfnamefont {I.}~\bibnamefont {Stavrakas}},
  \bibinfo {author} {\bibfnamefont {G.}~\bibnamefont {Hloupis}}, \ and\
  \bibinfo {author} {\bibfnamefont {D.}~\bibnamefont {Triantis}},\ }\href@noop
  {} {\bibfield  {journal} {\bibinfo  {journal} {Applied Physics B}\ }\textbf
  {\bibinfo {volume} {116}},\ \bibinfo {pages} {617} (\bibinfo {year}
  {2014})}\BibitemShut {NoStop}%
\end{thebibliography}
\end{document}